\documentclass[journal,10pt,twocolumn, cls]{IEEEtran} %
\usepackage{setspace}
\usepackage{multirow}
\usepackage{amsmath}
\usepackage{amsfonts}
\usepackage{cases}
\usepackage{fancybox}
\usepackage{subfigure}
\usepackage{epsfig}
\usepackage{graphicx}
\usepackage[justification=centering]{caption}
\usepackage{epstopdf}
\usepackage{float}
\usepackage{multirow}
\usepackage{dsfont}
\usepackage{times,amsmath,color,amssymb,epsfig,cite,subfigure,algorithm,algorithmic}

\newtheorem{theorem}{Theorem}

\begin{document}

\title{Optimized Signal Distortion for PAPR Reduction of OFDM Signals with IFFT/FFT Complexity via ADMM Approaches}

\author{Yongchao Wang, \IEEEmembership{Member, IEEE}, Yanjiao Wang and Qingjiang Shi}

\maketitle

\begin{abstract}
 \textcolor{black}{In this paper, we propose two low-complexity optimization methods to reduce peak-to-average power ratio (PAPR) values of orthogonal frequency division multiplexing (OFDM) signals via alternating direction method of multipliers (ADMM). First, we formulate a non-convex signal distortion optimization model based on minimizing data carrier distortion such that the constraints are placed on PAPR and the power of free carriers.
 Second, to obtain the model's approximate optimal solution efficiently, we design two low-complexity ADMM algorithms, named ADMM-Direct and ADMM-Relax respectively.
 Third, we show that, in ADMM-Direct/-Relax, all the optimization subproblems can be solved semi-analytically and the computational complexity in each iteration is roughly $\mathcal{O}(\ell N\log_2\ell N)$, where $\ell$ and $N$ are over-sampling factor and carrier number respectively.
 Moreover, we show that the resulting solution of ADMM-Direct is guaranteed to be some Karush-Kuhn-Tucker (KKT) point of the non-convex model when the iteration algorithm is convergent.
 For ADMM-Relax, we prove that it has theoretically guaranteed convergence and can approach arbitrarily close to some KKT point of the model if proper parameters are chosen.
 Simulation results demonstrate the effectiveness of the proposed approaches.}
\end{abstract}


\IEEEpeerreviewmaketitle

\section{Introduction}

\IEEEPARstart{O}RTHOGONAL frequency division multiplexing (OFDM) is an important multi-carrier modulation technique which has been used widely in modern wireless communication systems since it has high bandwidth efficiency and powerful ability to resist the effects of multi-path fading \cite{Bingham}.
However, a major drawback of OFDM signals is their high peak-to-average power ratio (PAPR). Since the transmitter's power amplifiers (PA) are peak-power limited, the large PAPR lets the wireless communication engineers face  a difficult dilemma between signal distortion and power efficiency \cite{Lee}.
The above dilemma can be seen from the following facts: On the one hand, to achieve high power amplifier efficiency, one can move working-point approaching to nonlinear region. Then, large signals would suffer from severe nonlinear distortion; On the other hand, to release nonlinear distortion of the large signals, one must move working-point back away from nonlinear region. Then, power efficiency would be low.

Over the past decades, there have been a variety of PAPR reduction techniques proposed in the literatures, which can be roughly classified into
three categories: multiple
signaling and probabilistic techniques, coding techniques, and signal distortion techniques \cite{Tutorial}.
The ideas of multiple signaling and probabilistic techniques, such as selective mapping (SLM) \cite{SLM}, partial transmit sequence (PTS) \cite{PTS}, tone reservation (TR) \cite{TR}, constellation shaping \cite{constellation_shaping}, etc., are used to generate multiple permutations of the OFDM signals and
transmit the one with a minimum PAPR, or to
modify the OFDM signals by introducing phase shifts, adding
peak reduction carriers, or changing constellation points to reduce the OFDM signals' PAPR.
The coding techniques use some coding schemes, for example low density parity-check (LDPC) code \cite{LDPC_code},  Hadamard code \cite{Hadamard_code}, etc., to perform PAPR reduction.
Signal distortion techniques reduce the PAPR by distorting the transmitted OFDM signal before it passes through the PA.
In comparison with other PAPR reduction techniques, signal distortion techniques have an important merit, which is that the signal distortion module can be inserted into the OFDM system directly and the corresponding transceiver's structure does not need to be changed.
Repeated clipping and filtering (RCF) \cite{clipping} may be the simplest signal distortion  method in the sense of computational complexity, which in every iteration is dominant by one fast fourier transform (FFT) operation and one inverse FFT (IFFT) operation.
However, on the one hand, the classical RCF technique and its variants, such as companding transform \cite{Hu_companding}, peak windowing \cite{peak_windowing}, peak cancellation \cite{peak_cancellation}, etc., cannot meet more complicated practical demands, such as controlling free carrier power under the specified level or achieving optimized signal distortion while approaching the desired PAPR values.
On the other hand,  as iteration algorithms, complete convergence analysis of these methods is still unavailable.

In recent years, signal distortion techniques based on optimization methods have been exploited to reduce the PAPR of OFDM signals while achieving optimal system parameters.
These kinds of optimization methods can make up the performance of the existing signal distortion methods, such as PAPR values and the corresponding signal distortion.
Second order conic programming (SOCP) approach was one of the widely used techniques, which is exploited to minimize peaks of the time-domain waveforms subject to constraints on error vector magnitude (EVM) and the free carrier power overhead (FCPO) \cite{SOCP}.
After that, several SOCP approaches were proposed to improve the PAPR performance of OFDM signals \cite{clipping_wang}-\!\!\cite{modified_SOICF}.
Semi-definite programming (SDP) is another important optimization technique to reduce the PAPR of the OFDM signals.
In \cite{SDR_Wang}, the authors exploited the semi-definite relaxation technique to relax the non-convex quadratic optimization model for OFDM signals and showed that the optimized OFDM symbols have a quasi-constant PAPR value.

The main concern of the existing PAPR optimization methods is their high-computational complexity.
In this paper, we focus on this issue and develop two low-complexity optimization methods via the alternating direction method of multipliers (ADMM) technique, whose complexities are comparable to the classical RCF method and also are calculated as dominant by one FFT operation and one IFFT operation.
The main content of this paper is as follows:
first, we establish a non-convex signal distortion optimization model which is based on minimizing data carrier distortion such that the constraints are placed on PAPR and the power of the free carriers.
To obtain approximate optimal solution of the non-convex model efficiently, we exploit the ADMM technique \cite{ADMM_Boyd}-\!\!\cite{Tom_Fast_ADMM}, and propose two customized ADMM algorithms, named ADMM-Direct and ADMM-Relax respectively.
It can be shown that, in both of the proposed algorithms, all the subproblems' optimal solutions can be determined semi-analytically and the computational complexity in each iteration is roughly $\mathcal{O}(\ell N\log_2\ell N)$, where $\ell$ and $N$ are the over-sampling factor and carrier number respectively.
Moreover, we show that the resulting solution of the ADMM-Direct algorithm is guaranteed to be some Karush-Kuhn-Tucker (KKT) point of the considered model when the iteration algorithm is convergent.
For ADMM-Relax, we prove that it is convergent and can approach arbitrarily close to some KKT point of the model if proper parameters are chosen.
Furthermore, the proposed ADMM algorithms outperform the existing approaches. For example, not only the desired OFDM symbols with quasi-constant PAPR values and small signal distortion can be obtained just after a few iterations but also convergence is theoretically guaranteed.

The rest of this paper is organized as follows. Section II introduces preliminaries related to OFDM signals and the considered OFDM optimization model. In Section III and IV, we exploit the ADMM technique and propose two low-complexity algorithms, named ADMM-Direct and ADMM-Relax respectively. Their performance analysis, such as convergence, convergence rate and complexity, are presented. Simulation results are shown to evaluate the performance of the proposed low-complexity OFDM PAPR reduction algorithms in Section V. Section VI concludes this paper.

\emph{Notations}: In this paper, bold lowercase and uppercase letters denote vectors and matrices respectively; $(\cdot)^T$ and $(\cdot)^H$ symbolize the transpose and conjugate transpose operations; 2-norm of a vector $\mathbf{a}$, $\infty$-norm of a vector $\mathbf{a}$ and the Frobenius norm of a matrix ${\bf A}$ are denoted by $||\mathbf{a}||_2$, $||\mathbf{a}||_\infty$ and $||\mathbf{A}||_{\rm F}$ respectively; $\nabla$ denotes the gradient operator. $\dagger$ denotes pseudo-inverse operator.

\section{Preliminaries}
Consider an OFDM system with $N$ carriers.
Let $\mathbf{c}\in\mathbb{C}^N$ denote an OFDM frequency-domain symbol and $\mathbf{x}\in \mathbb{C}^{\ell N}$ be its corresponding time-domain symbol.
Let $\mathbf{A}\in\mathbb{C}^{\ell N\times N}$ be the first $N$ columns of the $\ell N$-points IDFT matrix.
Then, there are
\begin{subequations}\label{IFFT}
\begin{align}
&\mathbf{x} = \mathbf{Ac} = \mathrm{IFFT}_{\ell}(\mathbf{c})\footnotemark, \\
&\mathbf{c} = \ell N\mathbf{A}^H\mathbf{x} = \mathrm{FFT}_{\ell}(\mathbf{x}),
\end{align}
\end{subequations}
where $\ell$ is the over-sampling factor, $\mathrm{IFFT}_{\ell}(\mathbf{c})$ denotes $\ell N$-points IFFT operation for the frequency-domain symbol $\mathbf{c}$ with $\ell$-times over-sampling, and $\mathrm{FFT}_{\ell}(\mathbf{x})$ denotes $\ell N$-points {\rm FFT} operation for the time-domain symbol $\mathbf{x}$, but only outputs the first $N$ elements.

\footnotetext{$\mathbf{x}$ can be cast as some discrete signal sampled from  continuous time domain OFDM signal.}

PAPR of the time-domain OFDM symbol $\mathbf{x}$ is defined as
\begin{equation}\label{PAPR2}
    {\rm{PAPR:}}~\frac{\max\limits_{i=1,\ldots,\ell N}|x_i|^2}
{\frac{1}{\ell N}\sum\limits_{i=1}^{\ell N}|x_i|^2}=\frac{\|\mathbf{x}\|_\infty^2}
{\frac{1}{\ell N}\|\mathbf{x}\|_2^2}.
\end{equation}
From \eqref{IFFT} and \eqref{PAPR2}, we see that there could exist large peaks in the time-domain OFDM symbol if carriers in the frequency-domain OFDM symbol are in phase or nearly in phase.

In many OFDM systems, carriers in an OFDM symbol consist of data carriers and free carriers.
The former are exploited to carry information and the latter are reserved to control the out-of-band emission or some future possible applications.
Generally, both introducing small distortion in the data carriers and assigning some controlled power to the free carriers can change PAPR values of the OFDM symbols. Based on these observations, we combine PAPR, data carrier distortion, free carrier power together and formulate the following optimization model
\begin{subequations}\label{ori model}
  \begin{align}
    &\underset{\displaystyle\mathbf{c}\in\mathbb{C}^N, \mathbf{x}\in\mathbb{C}^{\ell N}}{\min}\hspace{0.12cm} \frac{1}{2}\|\mathbf{S}_{\rm D}(\mathbf{c}-\mathbf{c}_{\rm o})\|_2^2,
              \label{ori model_a} \\
    &\hspace{0.55cm} {\rm subject\ to} \hspace{0.5cm} \frac{\|\mathbf{x}\|_\infty^2}
{\frac{1}{\ell N}\|\mathbf{x}\|_2^2} = \alpha,           \label{ori model_b} \\
    & \hspace{2.7cm} \displaystyle\frac{\|\mathbf{S}_{\rm F}\mathbf{c}\|_2^2}{\|\mathbf{S}_{\rm D}\mathbf{c}\|_2^2}\leq \beta, \label{ori model_c}   \\
    &  \hspace{2.7cm} \mathbf{Ac}=\mathbf{x}. \label{ori model_d}
  \end{align}
\end{subequations}
In the model \eqref{ori model}, $\mathbf{c}_{\rm o}$ is the original OFDM symbol.
The matrix $\mathbf{S}_{\rm D}$ is binary and diagonal.
The corresponding set ${\rm D}=\{i_m|m=1,\dotsb,M\}$ and $i_m$ labels the $m$th data carrier.
 $\mathbf{S}_{{\rm D}ii}=1$ if $i\in{\rm D}$ and $\mathbf{S}_{{\rm D}ii}=0$ otherwise.
The matrix $\mathbf{S}_{\rm F}$ and index set F have similar definitions except for the free carriers. The constraints \eqref{ori model_b} and \eqref{ori model_c} are PAPR constraint and free carriers constraint respectively, where $\alpha$ and $\beta$ are pre-set thresholds.

We have the following comments on the model \eqref{ori model}:
\begin{itemize}
\item Direct minimizing peak values of the OFDM time domain symbols is another optimization strategy. In comparison with it, the main benefit of the model \eqref{ori model} is that the optimized OFDM symbols have almost quasi-constant PAPR values, which can help us choose a proper working point of nonlinear PA.
\item According to the definitions of the matrices  $\mathbf{S}_{\rm D}$ and $\mathbf{S}_{\rm F}$, it can be seen that $\mathbf{S}_{\rm D}+\mathbf{S}_{\rm F}=\mathbf{I}$ and $\mathbf{S}_{\rm D}\mathbf{S}_{\rm F}=\mathbf{0}$, where $\mathbf{I}$ is an identity matrix.
\item To guarantee that the feasible region of the model \eqref{ori model} is non-empty, the pre-set thresholds $\alpha$ and $\beta$ should be set no less than 1 and 0 respectively, i.e., $\alpha\geq1$ and $\beta\geq0$.
\item The model \eqref{ori model} is designed to optimize the OFDM symbol whose PAPR is larger than $\alpha$.
    So, if the considered OFDM symbol's PAPR is less than $\alpha$, we do not process it and pass it to PA directly.
\item Since the constraints \eqref{ori model_b} and \eqref{ori model_c} are non-convex,
    it is difficult to obtain its global optimizer of the model \eqref{ori model}.
    Existing techniques, such as semi-definite relaxation, can be exploited to relax the model \eqref{ori model} to be convex and generate approximate optimal solutions.
    However, its computational cost is roughly $\mathcal{O}(\ell^{3.5}N^{3.5})$, which is prohibitive in practice.
    In the sequel, two low-complexity algorithms based on the ADMM technique for the model \eqref{ori model} are proposed and we show that their computational complexities in each iteration are roughly $\mathcal{O}(\ell N\log_2\ell N)$.
    Moreover, the desired OFDM symbol with quasi-constant PAPR values and optimized signal distortion can be obtained just after a few iterations and the proposed iteration algorithms have theoretically guaranteed convergence. In comparison with the proposed ADMM algorithms, the RCF method does not have these kinds of theoretical results.
\end{itemize}

\section{Solving algorithm I: ADMM-Direct}
ADMM is a simple but powerful technique that solves
large scale optimization problems by breaking them into small
ones, each of which is then easier to handle.
In this section, we propose the ADMM-Direct algorithm, which solves the problem \eqref{ori model} via the ADMM technique directly.
In ADMM-Direct, all the subproblems' optimal solutions can be determined  semi-analytically and the computational complexity in each iteration is roughly $\mathcal{O}(\ell N\log_2\ell N)$.
Moreover, we prove that the resulting solution of the ADMM-Direct algorithm is guaranteed to be some KKT point of the model \eqref{ori model} when the algorithm is convergent.

\subsection{ADMM-Direct Algorithm Framework}
The proposed ADMM-Direct algorithm is shown as follows.\begin{subequations}\label{ADMM Direct}
 \begin{align}
    {{\mathbf{c}}^{k+1}}&=\mathop{\arg \min }\limits_{\mathbf{c}\in\mathcal{C}}\  L_{\rho}(\mathbf{c},\mathbf{x}^k,\mathbf{y}^k), \label{DIR c_update}\\
    {{\mathbf{x}}^{k+1}}&=\mathop{\arg \min }\limits_{\mathbf{x}\in\mathcal{X}}\ L_{\rho}(\mathbf{c}^{k+1},\mathbf{x},\mathbf{y}^k), \label{DIR x_update}\\
    {{\mathbf{y}}^{k+1}}&={{\mathbf{y}}^{k}}+\rho(\mathbf{A}{{\mathbf{c}}^{k+1}}-{{\mathbf{x}}^{k+1}}). \label{DIR y_update}
 \end{align}
 \end{subequations}
 In \eqref{ADMM Direct}, $L_{\rho}(\mathbf{c},\mathbf{x},\mathbf{y})$ is the augmented Lagrangian function of the model \eqref{ori model} and it can be expressed as \footnotemark
 \begin{equation}\label{DIR ALG}
  \begin{split}
  L_{\rho}(\mathbf{c},\mathbf{x},\mathbf{y}) =& \frac{1}{2}\|\mathbf{S}_{\rm D}(\mathbf{c}-\mathbf{c}_{\rm o})\|_2^2 + {\rm Re}\big(\mathbf{y}^H(\mathbf{Ac}-\mathbf{x})\big)\\ &+\frac{\rho}{2}\|\mathbf{Ac}-\mathbf{x}\|_2^2,
  \end{split}
\end{equation}
where $\rho>0$ is the penalty parameter, $\mathbf{x}\in\mathcal{X}$ and $\mathbf{c}\in\mathcal{C}$ denote the constraints \eqref{ori model_b} and \eqref{ori model_c} respectively, $\mathbf{y}\in\mathbb{C}^{\ell N}$ is the Lagrangian multiplier, and $k$ is the iteration number.

\footnotetext{When the constraint is complex, one can introduce real Lagrangian multipliers $\mathbf{y}_{\rm R}$ and $\mathbf{y}_{\rm I}$ respectively for its real part and imaginary part. Then, according to the classical Lagrangian multiplier theory, the augmented Lagrangian function \eqref{DIR ALG} can be derived easily, where $\mathbf{y}=\mathbf{y}_{\rm R}+j\mathbf{y}_{\rm I}$.}

The challenges of implementing ADMM-Direct \eqref{ADMM Direct} are how to solve \eqref{DIR c_update} and \eqref{DIR x_update} since their corresponding constraints are non-convex.
In the following, we show that both of them can be obtained effectively by exploiting the structure of \eqref{ori model}.

\subsection{Solving the Subproblem \eqref{DIR c_update}}
Based on the augmented Lagrangian function $L_{\rho}(\mathbf{c},\mathbf{x},\mathbf{y})$, the problem \eqref{DIR c_update} can be equivalent to
{
\begin{subequations}\label{DIR c model}
  \begin{align}
    &\hspace{0.2cm}\underset{\displaystyle\mathbf{c}\in\mathbb{C}^N}{\min}\hspace{0.4cm} \frac{1}{2}\|\mathbf{S}_{\rm D}(\mathbf{c}-\mathbf{c}_{\rm o})\|_2^2 + \frac{\rho}{2}\|\mathbf{Ac}-\mathbf{x}^k + \frac{\mathbf{y}^k}{\rho}\|_2^2,             \label{DIR c model_a} \\
    &{\rm subject\ to} \hspace{0.2cm}\|\mathbf{S}_{\rm F}\mathbf{c}\|_2^2-\beta\|\mathbf{S}_{\rm D}\mathbf{c}\|_2^2\leq 0.   \label{DIR c model_b}
  \end{align}
\end{subequations}}
Since there is only one constraint in \eqref{DIR c model}, its optimal solution can be determined through the Lagrangian multiplier method.
The Lagrangian function of model \eqref{DIR c model} can be written as
\begin{equation}\label{DIR lagrangian c}
 \begin{split}
  L(\mathbf{c},\mu^k) &= \frac{1}{2}\|\mathbf{S}_{\rm D}(\mathbf{c}-\mathbf{c}_{\rm o})\|_2^2 + \frac{\rho}{2}\|\mathbf{Ac}-\mathbf{x}^k + \frac{\mathbf{y}^k}{\rho}\|_2^2 \\ &\hspace{0.5cm}+ \mu^k\big(\|\mathbf{S}_{\rm F}\mathbf{c}\|_2^2-\beta\|\mathbf{S}_{\rm D}\mathbf{c}\|_2^2\big),
  \end{split}
\end{equation}
where the Lagrangian multiplier is $\mu^k\geq0$.
Since the problem \eqref{DIR c model} is feasible, the Lagrangian multiplier theorem indicates that its global optimal solution $\mathbf{c}^{k+1}$, combining the optimal Lagrangian multiplier $\mu^{k*}$, should satisfy $\displaystyle\nabla_\mathbf{c} L(\mathbf{c}^{k+1},\mu^{k*})=0$, i.e.,
\begin{equation}\label{gradient c1}
   \begin{split}
   \hspace{-0.2cm}\nabla_\mathbf{c} L(\!\mathbf{c}^{k+1}\!,\mu^{k*}\!) \!&=\!\mathbf{S_{\rm D}}\!(\mathbf{c}^{k+1}\!-\!\mathbf{c}_{\rm o}\!)\!+\! \rho\mathbf{A}^H\!\big(\!\mathbf{A}\mathbf{c}^{k+1}\!-\!\mathbf{x}^k\!+\!\frac{\mathbf{y}^k}{\rho}\!\big) \\ &\hspace{0.5cm}+ 2\mu^{k*}(\mathbf{S}_{\rm F}-\beta\mathbf{S}_{\rm D})\mathbf{c}^{k+1} = 0.
   \end{split}
\end{equation}
Since $\mathbf{A}^H\mathbf{A}=\frac{\mathbf{I}}{\ell N}$ and $\mathbf{c}_{\rm o}=\mathbf{S}_{\rm D}\mathbf{c}_{\rm o}$, we can further derive \eqref{gradient c1} as
\[
  \big(\mathbf{S}_{\rm D}+\frac{\rho}{\ell N}\mathbf{I}\! +\! 2\mu^{k*}(\mathbf{S}_{\rm F}-\beta\mathbf{S}_{\rm D})\big)\mathbf{c}^{k+1}= \mathbf{c}_{\rm o}+\!\rho\mathbf{A}^H(\mathbf{x}^k-\!\frac{\mathbf{y}^k}{\rho}).
\]
Then, we can obtain
\begin{equation}\label{DIR c opt mu}
  \mathbf{c}^{k+1} = \big(\mathbf{S}_{\rm D}+\frac{\rho}{\ell N}\mathbf{I} + 2\mu^{k*}(\mathbf{S}_{\rm F}-\beta\mathbf{S}_{\rm D})\big)^{\dagger}\mathbf{v}^k,
\end{equation}
where $\displaystyle\mathbf{v}^k=\mathbf{c}_{\rm o}+\rho\mathbf{A}^H(\mathbf{x}^k-\frac{\mathbf{y}^k}{\rho})$.

Now we consider how to determine $\mu^{k*}$. If the constraint \eqref{DIR c model_b} is inactive, the corresponding Lagrangian multiplier $\mu^{k*}=0$.
Otherwise, if the constraint \eqref{DIR c model_b} is active, $\mathbf{c}^{k+1}$ should satisfy the constraint \eqref{DIR c model_b} when ``='' holds. We first consider the latter. Plugging \eqref{DIR c opt mu} into $\|\mathbf{S}_{\rm F}\mathbf{c}^{k+1}\|_2$ and $\|\!\mathbf{S}_{\rm D}\mathbf{c}^{k+1}\!\|_2$, we have
\begin{subequations}
    \begin{align}
        \|\!\mathbf{S}_{\rm F}\mathbf{c}^{k+1}\!\|_2\!&= \frac{\|\mathbf{S}_{\rm F}\mathbf{v}^k\|_2}{\frac{\rho}{\ell N} + 2\mu^{k*}}, \label{DIR Sf} \\
        \|\!\mathbf{S}_{\rm D}\mathbf{c}^{k+1}\!\|_2\!&= \frac{\|\mathbf{S}_{\rm D}\mathbf{v}^k\|_2}{1+\frac{\rho}{\ell N} - 2\mu^{k*}\beta}. \label{DIR Sd}
    \end{align}
\end{subequations}
Notice here we use the properties that $\mathbf{S}_{\rm D}$ and $\mathbf{S}_{\rm F}$ are diagonal matrices and $\mathbf{S}_{\rm D}\mathbf{S}_{\rm F}=0$.
 Plugging \eqref{DIR Sf} and \eqref{DIR Sd} into $\|\mathbf{S}_{\rm F}\mathbf{c}^{k+1}\|_2^2=\beta\|\mathbf{S}_{\rm D}\mathbf{c}^{k+1}\|_2^2$, we obtain
\begin{equation}\label{DIR mu_k}
  \mu^{k*}\! =\! \frac{\big(1+\frac{\rho}{\ell N}\big)\|\mathbf{S}_{\rm F}\mathbf{v}^k\|_2-\sqrt{\beta}\frac{\rho}{\ell N}\|\mathbf{S}_{\rm D}\mathbf{v}^k\|_2}{2(\beta\|\mathbf{S}_{\rm F}\mathbf{v}^k\|_2+\sqrt{\beta}\|\mathbf{S}_{\rm D}\mathbf{v}^k\|_2)}.
\end{equation}

In the implementation, we still need to know in what case we use \eqref{DIR mu_k} to compute $\mu^{k*}$ or just set $\mu^{k*}$ as zero.
Observing \eqref{DIR mu_k}, we see that the computed result for $\mu^{k*}$ could be negative.
However, the Lagrangian multiplier theory guarantees that $\mu^{k*}$ should always be nonnegative since it is for the inequality constraint \eqref{DIR c model_b}.
This contradiction comes from the assumption that the constraint is active.
It means that the constraint is inactive and so $\mu^{k*}$ should be zero.
Based on this observation, we can compute $\mu^{k*}$ by
\begin{equation}\label{DIR mu_c}
\!\mu^{k*}\!=\!\max\!\bigg\{\!0,\!\frac{\big(1\!+\!\frac{\rho}{\ell N}\big)\|\mathbf{S}_{\rm F}\mathbf{v}^k\|_2\!-\!\sqrt{\beta}\frac{\rho}{\ell N}\|\mathbf{S}_{\rm D}\mathbf{v}^k\|_2}{2(\beta\|\mathbf{S}_{\rm F}\mathbf{v}^k\|_2+\sqrt{\beta}\|\mathbf{S}_{\rm D}\mathbf{v}^k\|_2)}\bigg\}.
\end{equation}

\begin{table}[htbp]
\renewcommand \arraystretch{1.0}
\centering
\caption{Binary section searching procedure for $\gamma^{k*}$}
\begin{tabular}{l}
\hline \hline
\textbf{Initialization:} Set search boundary ($\gamma^k_{\rm left}$, $\gamma^k_{\rm right}$). To\\
guarantee $\gamma^{k*}\in(\gamma^k_{\rm left}$, $\gamma^k_{\rm right}$), we set
$\gamma^k_{\rm left}=0$
 and $\gamma^k_{\rm right}$\\ is large enough. \\
\textbf{Repeat:} Let $\displaystyle\gamma^k=\frac{\gamma^k_{\rm left}+\gamma^k_{\rm right}}{2}$. Update $\mathbf{z}^{k+1}$ using \eqref{DIR zi}. \\
If $\|\mathbf{z}^{k+1}\|_2^2<1$, set $\gamma^k_{\rm right}=\gamma^k $.
Otherwise set $\gamma^k_{\rm left}=\gamma^k$. \\
\textbf{Until} $\|\mathbf{z}^{k+1}\|_2^2$ is close to 1 enough and let\\
 $\gamma^{k*}=\frac{\displaystyle\gamma^k_{\rm left}+\gamma^k_{\rm right}}{2}$. \\
\hline \hline
\end{tabular}
\label{bisection}
\end{table}

\subsection{Solving the Subproblem \eqref{DIR x_update}}
Based on the augmented Lagrangian function $L_{\rho}(\mathbf{c},\mathbf{x},\mathbf{y})$, the subproblem \eqref{DIR x_update} can be equivalent to
\begin{subequations}\label{DIR x model}
  \begin{align}
    &\hspace{0.1cm}\underset{\displaystyle\mathbf{x}\in\mathbb{C}^{\ell N}}{\min}\hspace{0.5cm} \|\mathbf{Ac}^{k+1}-\mathbf{x} + \frac{\mathbf{y}^k}{\rho}\|_2^2,             \label{DIR x model_a} \\
    &{\rm subject\ to} \hspace{0.5cm} \frac{\|\mathbf{x}\|_\infty^2}
{\frac{1}{\ell N}\|\mathbf{x}\|_2^2} = \alpha .           \label{DIR x model_b}
  \end{align}
\end{subequations}
To simplify the constraint \eqref{DIR x model_b}, we introduce auxiliary variables $t$ and $\mathbf{z}$ to express $\mathbf{x}$ by $\mathbf{x}=t\mathbf{z}$,
where $t>0$ and $\|\mathbf{z}\|_2^2=1$. Plugging them into \eqref{DIR x model}, it is equivalent to
\begin{subequations}\label{DIR x model t}
  \begin{align}
    &\underset{\displaystyle\mathbf{z}\in\mathbb{C}^{\ell N}, t>0}{\min}\hspace{0.3cm} t^2-2t{\rm Re}(\mathbf{z}^H\mathbf{b}^k),             \label{DIR x model t_a} \\
    &\hspace{0.4cm}{\rm subject\ to} \hspace{0.58cm} |z_i|^2 \leq \frac{\alpha}{\ell N}, \ \ i=1,\dotsb,\ell N,   \label{DIR x model t_b} \\
    &\hspace{2.5cm}  \|\mathbf{z}\|_2^2 = 1, \label{DIR x model t_c}
  \end{align}
\end{subequations}
where
$
\displaystyle\mathbf{b}^k=\mathbf{Ac}^{k+1}+\frac{\mathbf{y}^k}{\rho}
$
and $z_i\in\mathbf{z}$.

  To be clear, we use $\mathbf{z}^{k+1}$ and $t^{k+1}$ to denote the optimal solutions of the model \eqref{DIR x model t}.
  Apparently, to minimize the objective \eqref{DIR x model t_a}, ${\rm Re}(\mathbf{z}^H\mathbf{b}^k)$ should be maximized subject to \eqref{DIR x model t_b} and \eqref{DIR x model t_c}.
  Specifically, we drop $t$ from the model \eqref{DIR x model t} and formulate \eqref{DIR x model z t} to solve $\mathbf{z}^{k+1}$.
\begin{subequations}\label{DIR x model z t}
  \begin{align}
    &\underset{\displaystyle\mathbf{z}\in\mathbb{C}^{\ell N}}{\rm max}\hspace{0.6cm} {\rm Re}(\mathbf{z}^H\mathbf{b}^k),             \label{DIR x model z t_a} \\
    &{\rm subject\ to} \hspace{0.3cm} |z_i|^2 \leq \frac{\alpha}{\ell N}, \ \ i=1,\dotsb,\ell N,    \label{DIR x model z_b} \\
    &\hspace{1.85cm}  \|\mathbf{z}\|_2^2 = 1. \label{DIR x model z t_c}
  \end{align}
\end{subequations}
Moreover, we can further change \eqref{DIR x model z t_c} to an inequality constraint and formulate an equivalent convex optimization model \eqref{DIR x model z convex}, which can be solved as an SOCP problem with computational complexity $\mathcal{O}(\ell^3 N^3 )$ using free optimization solver \cite{sedumi}\cite{CVX}. Here, we say that the models \eqref{DIR x model z t} and \eqref{DIR x model z convex} are equivalent in the sense that both of them have the same optimal solution. We prove this fact in Appendix A.
\begin{subequations}\label{DIR x model z convex}
  \begin{align}
    &\underset{\displaystyle\mathbf{z}\in\mathbb{C}^{\ell N}}{\rm max}\hspace{0.6cm} {\rm Re}(\mathbf{z}^H\mathbf{b}^k),             \label{DIR x model convex z_a} \\
    &{\rm subject\ to} \hspace{0.3cm} |z_i|^2 \leq \frac{\alpha}{\ell N}, \ \ i=1,\dotsb,\ell N,   \label{DIR x model convex z_b} \\
    &\hspace{1.9cm}  \|\mathbf{z}\|_2^2 \leq 1. \label{DIR x model convex z_c}
  \end{align}
\end{subequations}
From a practical viewpoint, solving the problem \eqref{DIR x model z convex} with complexity $\mathcal{O}(\ell^3 N^3 )$ is still expensive.
In the following, we devise an inexact parallel solving algorithm, which can be implemented very effectively.
First, we introduce the Lagrangian multiplier $\gamma^k>0$ for the constraint \eqref{DIR x model convex z_c} and rewrite \eqref{DIR x model z convex} as
\begin{subequations}\label{DIR x model z sep}
  \begin{align}
    &\underset{\displaystyle{z_i}\in\mathbb{C},\! \gamma^k>0}{\min}\hspace{0.35cm} \displaystyle\sum_{i=1}^{\ell N}\!-{\rm Re}(z_i^\dagger b_i^k)\!+\!{\gamma}^k\!\bigg(\!\displaystyle\sum_{i=1}^{\ell N}|z_i|^2\!-\!1\!\bigg)\!,\label{DIR x model z_spe a} \\
    &\hspace{0.38cm} {\rm subject\ to} \hspace{0.7cm} |z_i| \leq \sqrt{\frac{\alpha}{\ell N}}, \ \ i=1,\dotsb,\ell N,   \label{DIR x model z_spe b}
  \end{align}
\end{subequations}
where $``\dagger''$ denotes the conjugate operator.
Since both the objective function \eqref{DIR x model z_spe a} and constraint \eqref{DIR x model z_spe b} can be treated separately in $z_i$, solving the model \eqref{DIR x model z sep} is equivalent to solving the following $\ell N$ subproblems \eqref{DIR x model z single}, which can be implemented in parallel.
\begin{subequations}\label{DIR x model z single}
  \begin{align}
    &\underset{\displaystyle{z_i}\in\mathbb{C}, \gamma^k>0}{\min}\hspace{0.3cm} -{\rm Re}(z_i^\dagger b_i^k) + \gamma^k|z_i|^2,    \label{DIR x model z_single a} \\
    &\hspace{0.48cm} {\rm subject\ to} \hspace{0.6cm} |z_i| \leq \sqrt{\frac{\alpha}{\ell N}}.   \label{DIR x model z_single b}
  \end{align}
\end{subequations}
Since \eqref{DIR x model z_single a} is a convex quadratic function and the constraint \eqref{DIR x model z_single b} involves only one variable, the model's optimizer can be obtained by setting the objective's gradient as zero and then projecting the corresponding equation's solution onto the feasible region, i.e.,
\begin{equation}\label{DIR zi}
  z_i^{k+1} = \begin{cases} \hspace{0.5cm} \displaystyle\frac{b_i^k}{2\gamma^k}, \hspace{1cm} \frac{|b_i^k|}{2\gamma^k}<\sqrt\frac{\alpha}{\ell N}, \\
  \sqrt\frac{\alpha}{\ell N}e^{\displaystyle j\phi(b_i^k)}, \ \ {\rm otherwise},
  \end{cases}
\end{equation}
where $\phi(b_i^k)$ represents the phase of $b_i^k$.
Moreover, the optimal Lagrangian multiplier $\gamma^{k*}$ can be obtained effectively through the binary section searching procedure as shown in Table \ref{bisection}.
Then, plugging $\mathbf{z}^{k+1}$ into the model \eqref{DIR x model t} and simplifying it as a quadratic problem, we can get
\begin{equation}
t^{k+1}={\rm Re}\big(\mathbf{z}^{k+1 H}\mathbf{b}^k\big).
\end{equation}
According to \eqref{DIR zi}, we can find that ${\rm Re}\big(\mathbf{z}^{k+1 H}\mathbf{b}^k\big)$ is guaranteed to be positive. Thus, the constraint $t>0$ is always satisfied.
Plugging $\mathbf{z}^{k+1}$ and $t^{k+1}$ into $\mathbf{x}=t\mathbf{z}$, we get $\mathbf{x}^{k+1}$.

\begin{figure}[htbp]
\setlength{\abovecaptionskip}{-0.3cm}
\setlength{\belowcaptionskip}{-0.4cm}
    \begin{center}
    \begin{spacing}{1.1}
    \ovalbox{
     \begin{Bflushleft}[b]
       \ \ \textbf{Initialization}: Initialize ($\mathbf{c}^1, \mathbf{x}^1, \mathbf{y}^1$). Choose parameters\\
        ($\alpha, \beta, \rho$).
       Based on the considered OFDM scheme, set\\
        diagonal matrices $\mathbf{S}_{\rm D}$ and $\mathbf{S}_{\rm F}$.\\
             \textbf{For} $k=1,\ 2,\ 3 \dotsb$  \\
               \hspace{0.1cm} S.1 \ Solve the subproblem \eqref{DIR c_update}. \\
               \hspace{0.3cm} 1.1 \ Compute $\displaystyle\mathbf{v}^k=\mathbf{c}_{\rm o}+\frac{\rho}{\ell N}{\rm FFT}_{\ell}(\mathbf{x}^k-\frac{\mathbf{y}^k}{\rho})$.
               \\
               \hspace{0.3cm} 1.2 \ Compute \\
               \hspace{0.5cm}$\mu^{k*}\! = \!\max\!\bigg\{\!0,\!\displaystyle\frac{\big(\!1\!+\!\frac{\rho}{\ell N}\!\big)\|\mathbf{S}_{\rm F}\mathbf{v}^k\|_2\!-\!\sqrt{\beta}\frac{\rho}{\ell N}\|\mathbf{S}_{\rm
               D}\mathbf{v}^k\|_2}{2(\beta\|\mathbf{S}_{\rm F}\mathbf{v}^k\|_2\!+\!\sqrt{\beta}\|\mathbf{S}_{\rm D}\mathbf{v}^k\|_2)}\!\bigg\}$. \\
               \hspace{0.3cm} 1.3 \ Compute \\
               \hspace{0.5cm}$\mathbf{c}^{k+1} = \big(\mathbf{S}_{\rm D}+\frac{\rho}{\ell N}\mathbf{I} + 2\mu^{k*}(\mathbf{S}_{\rm F}-\beta\mathbf{S}_{\rm D})\big)^{\dagger}\mathbf{v}^k$. \\
               \hspace{0.1cm} S.2 \ Solve the subproblem \eqref{DIR x_update}. \\
               \hspace{0.3cm} 2.1 \ Compute $\displaystyle\mathbf{b}^k={\rm IFFT}_{\ell}(\mathbf{c}^{k+1})+\frac{\mathbf{y}^k}{\rho}$. \\
               \hspace{0.3cm} 2.2 \ Compute $\mathbf{z}^{k+1}$ through the binary section\\
                \hspace{0.5cm}searching procedure in Table \ref{bisection}. \\
               \hspace{0.3cm} 2.3 \ Compute $t^{k+1}=\max\{0, {\rm Re}\big(\mathbf{z}^{k+1 H}\mathbf{b}^k\big)\}$. \\
               \hspace{0.3cm} 2.4 \ Compute $\mathbf{x}^{k+1} = t^{k+1}\mathbf{z}^{k+1}$. \\
               \hspace{0.1cm} S.3 \ Update Lagrangian multipliers. \\
               \hspace{0.3cm}  Compute ${{\mathbf{y}}^{k+1}}=\mathbf{y}^k + \rho({\rm IFFT}_{\ell}(\mathbf{c}^{k+1})-\mathbf{x}^{k+1})$. \\
               \textbf{\ Until} some preset termination conditions are satisfied.\\
               \hspace{0.9cm} Then, let $\mathbf{x}^{k+1}$ be the output.
     \end{Bflushleft}
     }
     \end{spacing}
     \end{center}
     \caption{ADMM-Direct algorithm for the model \eqref{ori model}.}
     \label{ADMM DIR}
\end{figure}

\subsection{Performance Analysis on the ADMM-Direct Algorithm}
\subsubsection{Computational complexity} In each ADMM-Direct iteration, the computational cost is quite cheap, which is comparable to the classical RCF approach \cite{clipping}.
For ADMM-Direct algorithm scheme in Figure 1, we first consider the computational complexity of solving $\mathbf{c}^{k+1}$.
In S.1.1, when we compute $\mathbf{v}^{k}$, it is obvious that the main cost lies in computing ${\rm FFT}_{\ell}(\mathbf{x}^k-\frac{\mathbf{y}^k}{\rho})$.
Since $\mathbf{x}^k-\frac{\mathbf{y}^k}{\rho}$ is an $\ell N$-length vector, the computational complexity to determine $\mathbf{v}^{k}$ is roughly $\mathcal{O}(\ell N \log_2\ell N)$.
Notice here that $\ell N$ points {\rm FFT} operation can be implemented through $\frac{1}{2}\ell N \log_2\ell N$ complex multiplications. In S.1.2, since $\mathbf{S}_{\rm D}$ and $\mathbf{S}_{\rm F}$ are binary and diagonal matrices,
 it costs only $\mathcal{O}(N)$ complex multiplications to obtain $\|\mathbf{S}_{\rm D}\mathbf{v}^k\|_2$ and $\|\mathbf{S}_{\rm F}\mathbf{v}^k\|_2$, i.e., the computational cost to determine $\mu^{k*}$ is roughly $\mathcal{O}(2N)$.
 In S.1.3, since the matrix $\mathbf{S}_{\rm D}+\frac{\rho}{\ell N}\mathbf{I} + 2\mu^{k*}(\mathbf{S}_{\rm F}-\beta\mathbf{S}_{\rm D})$ is diagonal, its pseudo-inverse can be implemented using $N$ complex multiplications. Summarizing S.1.1-S.1.3, we can conclude that the computational cost to determiner $\mathbf{c}^{k+1}$ is dominant by $\ell N$-points {\rm IFFT} operation, i.e., roughly $\mathcal{O}(\ell N \log_2\ell N)$.
 Second, we analyze the computational cost to obtain $\mathbf{x}^{k+1}$. In S.2.1, the main cost to compute $\mathbf{b}^k$ lies in ${\rm IFFT}_{\ell}(\mathbf{c}^{k+1})$. Since implementing $\ell N$ points {\rm IFFT} operation needs $\frac{1}{2}\ell N \log_2\ell N$ complex multiplications, we can obtain $\mathbf{b}^k$ through roughly $\mathcal{O}(\ell N \log_2\ell N)$ complex multiplications. In S.2.2,  Bi-section searching procedure is exploited to determine $\mathbf{z}^{k+1}$. Observing (19), we can see that every element $z_i^{k+1}$ in $\mathbf{z}^{k+1}$ can be obtained just through one multiplication. Notice here $\sqrt{\frac{\alpha}{\ell N}}$ is constant and can be reused for computing all elements in $\mathbf{z}^{k+1}$.
 Here, it should note that the solution accuracy of the Bi-section searching procedure depends on the iteration number and the value $\gamma_{\rm right}^k$ (see Table I).
 Usually, when it takes several iterations, for example 10, and the corresponding $\gamma_{\rm right}^k=100$ , pretty good solution, for example one percent accuracy in {\rm PAPR dB} can be obtained, which is enough for the practical applications. Therefore, the computational complexity to obtain $\mathbf{z}^{k+1}$ is comparable to or less than implementing ${\rm IFFT}_{\ell}(\mathbf{c}^{k+1})$, especially when $\ell$ and $N$ are large.
 Since the costs of implementing S.2.3 and S.2.4 are far less than that of S.2.1 and S.2.2, it also takes roughly $\mathcal{O}(\ell N \log_2\ell N)$ complex multiplications to compute $\mathbf{x}^{k+1}$.
 At last, in S.3, since ${\rm IFFT}_{\ell}(\mathbf{c}^{k+1})$  is already obtained in S.2.1, $\mathbf{y}^{k+1}$ can be obtained through $\ell N$ complex multiplications.
 Combining the above analysis on S.1-S.3, we can conclude that the total computational cost in each ADMM-Direct iteration is the order  $\mathcal{O}(\ell N\log_2\ell N)$.

Moreover, we should note that it may take many iterations to let ADMM-Direct converge, which could be a significant burden in practice. However, in the simulation section, we show that good OFDM symbol, i.e., with quasi-constant PAPR and very small distortion, can be obtained just after a few iterations.

\subsubsection{Convergence issue} We have the following theorem on ADMM-Direct algorithm. Its proof is shown in Appendix B.
\begin{theorem}\label{convergence ADMM DIR}
    Let $\{\mathbf{c}^k, \mathbf{x}^k, \mathbf{y}^k, k=1,2,\dotsb\}$ be the tuples generated by the proposed ADMM-Direct algorithm. If $\underset{k\rightarrow+\infty}\lim\{\mathbf{c}^k, \mathbf{x}^k$, $\mathbf{y}^k\}=(\mathbf{c}^*, \mathbf{x}^*, \mathbf{y}^*)$, then $(\mathbf{c}^*, \mathbf{x}^*, \mathbf{y}^*)$ is some KKT point of the model \eqref{ori model}.
\end{theorem}

{\it Remarks:}
Here, we should note that the above Theorem \ref{convergence ADMM DIR} just states the quality of the solution when the ADMM-Direct algorithm is convergent.
Exact convergence analysis is difficult since the feasible region in the model \eqref{ori model} is non-convex.
Actually, the convergence analysis of the ADMM algorithm for the general non-convex optimization problem is still open to date. Existing analysis methods, such as in \cite{Yin_Wo_tao} and \cite{Nonconvex_ADMM_Hong}, cannot be exploited since the non-convex model \eqref{ori model} cannot satisfy their specifical conditions.
However, the simulation results in this paper show that the proposed ADMM-Direct algorithm always converges, and the resulting OFDM symbols have good practical performance, i.e. quasi-constant PAPR values.
In the next section, we develop a different ADMM algorithm named ADMM-Relax for the model \eqref{ori model}, which is theoretically guaranteed to be convergent and can be arbitrarily close to some KKT point of the model \eqref{ori model} if proper penalty parameters are chosen.

\section{Solving Algorithm II: ADMM-Relax}
In this section, we propose the ADMM-Relax algorithm. In ADMM-Relax, we relax the model \eqref{ori model} to the model \eqref{relax model} and then we use the ADMM technique to solve the model \eqref{relax model}.
In this algorithm, all the subproblems' optimal solutions can be determined  semi-analytically and the computational complexity in each iteration is roughly $\mathcal{O}(\ell N\log_2\ell N)$.
Furthermore, we prove that ADMM-Relax is convergent and can approach arbitrarily close to some KKT point of the model \eqref{ori model} if proper parameters are chosen. Morever, the simulation results show that the optimal OFDM symbols optimized by ADMM-Relax through a few iterations have quasi-constant PAPR values.
\subsection{ADMM-Relax Algorithm Framework}
In ADMM-Relax, we relax the model \eqref{ori model} to the model \eqref{relax model} by introducing auxiliary variables $\mathbf{u}$ and $\mathbf{w}$ for the constraint \eqref{ori model_d} and adding penalty  $\|\mathbf{u}-\mathbf{w}\|_2^2$ to the objective function. The proposed ADMM-Relax algorithm is shown as follows
\begin{subequations}\label{relax model}
  \begin{align}
    &\underset{\mathbf{c}\in\mathbb{C}^N, \mathbf{x},\mathbf{u,w}\in\mathbb{C}^{\ell N}}{\min}\hspace{0.2cm} \frac{1}{2}\|\mathbf{S}_{\rm D}(\mathbf{c}-\mathbf{c}_{\rm o})\|_2^2 + \frac{\tilde{\rho}}{2}\|\mathbf{u}-\mathbf{w}\|_2^2,
              \label{relax model_a} \\
    &\hspace{0.55cm} {\rm subject\ to} \hspace{0.5cm} \frac{\|\mathbf{x}\|_\infty^2}
{\frac{1}{\ell N}\|\mathbf{x}\|_2^2} = \alpha,           \label{relax model_b} \\
    & \hspace{2.7cm} \displaystyle\frac{\|\mathbf{S}_{\rm F}\mathbf{c}\|_2^2}{\|\mathbf{S}_{\rm D}\mathbf{c}\|_2^2}\leq \beta, \label{relax model_c}   \\
    &  \hspace{2.7cm} \mathbf{Ac}=\mathbf{u}, \label{relax model_d} \\
    &  \hspace{2.7cm} \mathbf{x}=\mathbf{w}, \label{relax model_e}
  \end{align}
\end{subequations}
where $\tilde{\rho}>0$ is the penalty factor. Intuitively, $\mathbf{u}$ and $\mathbf{w}$ can be arbitrarily close if $\tilde{\rho}$ is large enough. The augmented Lagrangian function for the model \eqref{relax model} is formulated as
 \begin{equation}\label{Relax ALG}
 \begin{split}
  L_{\rho}(\mathbf{c},\mathbf{x},\mathbf{u},&\mathbf{w},\mathbf{y}_1,\mathbf{y}_2) = \frac{1}{2}\|\mathbf{S}_{\rm D}(\mathbf{c}-\mathbf{c}_{\rm o})\|_2^2 \\
  +& {\rm Re}\big(\mathbf{y}_1^H(\mathbf{Ac}-\mathbf{u})\big)+{\rm Re}\big(\mathbf{y}_2^H(\mathbf{x}-\mathbf{w})\big) \\ +&\frac{\tilde{\rho}}{2}\|\mathbf{u}-\mathbf{w}\|_2^2+
  \frac{\rho}{2}(\|\mathbf{Ac}-\mathbf{u}\|_2^2+\|\mathbf{x}-\mathbf{w}\|_2^2),
 \end{split}
\end{equation}
 where $\mathbf{y}_1\in\mathbb{C}^{\ell N}$ and $\mathbf{y}_2\in\mathbb{C}^{\ell N}$ are Lagrangian multipliers corresponding to the constraints \eqref{relax model_d} and \eqref{relax model_e} respectively.
The proposed ADMM-Relax algorithm for the model \eqref{ori model} is formulated as follows
 \begin{subequations}\label{Relax ADMM iteration}
 \begin{align}
    &{{\mathbf{c}}^{k+1}}=\mathop{\arg \min }\limits_{\mathbf{c}\in\mathcal{C}}\  L_{\rho}(\mathbf{c},\mathbf{x}^k, \mathbf{u}^k, \mathbf{w}^k, \mathbf{y}_1^k,  \mathbf{y}_2^k), \label{Relax c_update}\\
    &{{\mathbf{x}}^{k+1}}=\mathop{\arg \min }\limits_{\mathbf{x}\in\mathcal{X}}\ L_{\rho}(\mathbf{c}^{k+1},\mathbf{x}, \mathbf{u}^k, \mathbf{w}^k, \mathbf{y}_1^k,  \mathbf{y}_2^k), \label{Relax x_update}\\
    &(\!{{\mathbf{u}}^{k+1},\mathbf{w}^{k+1}}\!)\!= \!\mathop{\arg \min}\limits_{\mathbf{u,w}\in\mathbb{C}^{\ell N}}
     L_{\rho}\!(\!\mathbf{c}^{k+1},\mathbf{x}^{k+1}, \mathbf{u}, \mathbf{w}, \mathbf{y}_1^k,  \mathbf{y}_2^k\!), \label{Relax u_update}\\
    &{{\mathbf{y}}_1^{k+1}}={{\mathbf{y}}_1^{k}}+\rho(\mathbf{A}{{\mathbf{c}}^{k+1}}-{{\mathbf{u}}^{k+1}}), \label{Relax y1_update} \\
     &{{\mathbf{y}}_2^{k+1}}={{\mathbf{y}}_2^{k}}+\rho(\mathbf{x}^{k+1}-{{\mathbf{w}}^{k+1}}), \label{Relax y2_update}
 \end{align}
 \end{subequations}
 where $\mathbf{x}\in\mathcal{X}$ and $\mathbf{c}\in\mathcal{C}$ denote the constraints \eqref{relax model_b} and \eqref{relax model_c} respectively, and $k$ is the iteration number.
 Solving \eqref{Relax c_update} and \eqref{Relax x_update} are quite similar to solving \eqref{DIR c_update} and \eqref{DIR x_update}.
 Their optimal solutions can also be determined semi-analytically.
Moreover, \eqref{Relax u_update} is an unconstrained convex quadratic problem. It means that its optimal solutions can also be expressed in close-form. Detailed derivations for ($\mathbf{c}^{k+1},\mathbf{x}^{k+1}, \mathbf{u}^{k+1}, \mathbf{w}^{k+1}$) are presented in Appendix C. In Figure \ref{ADMM-Relax}, we summarize the proposed ADMM-Relax algorithm for the model \eqref{ori model}.

\begin{figure}[htbp]
\setlength{\abovecaptionskip}{-0.3cm}
\setlength{\belowcaptionskip}{-0.5cm}
    \begin{center}
    \begin{spacing}{1.1}
    \ovalbox{
     \begin{Bflushleft}[b]
       \ \ \textbf{Initialization}: Initialize ($\mathbf{c}^1, \mathbf{x}^1$, $\mathbf{u}^1$, $\mathbf{w}^1$, $\mathbf{y}_1^1$, $\mathbf{y}_2^1$).\\
        Choose parameters ($\alpha, \beta, \rho, \tilde{\rho}$). Based on the considered\\
         OFDM scheme, set diagonal matrices $\mathbf{S}_{\rm D}$ and $\mathbf{S}_{\rm F}$.\\
             \textbf{For} $k=1,\ 2,\ 3 \dotsb$  \\
               \hspace{0.1cm} S.1 \ Solve the subproblem \eqref{Relax c_update}. \\
               \hspace{0.3cm} 1.1 \ Compute $\displaystyle\mathbf{v}^k=\mathbf{c}_{\rm o}+\frac{\rho}{\ell N}{\rm FFT}_{\ell}(\mathbf{u}^k-\frac{\mathbf{y}_1^k}{\rho})$.
               \\
               \hspace{0.3cm} 1.2 \ Compute\\
                \hspace{0.5cm}$\mu^{k*}\! = \!\max\!\bigg\{\!0,\!\displaystyle\frac{\big(\!1\!+\!\frac{\rho}{\ell N}\!\big)\|\mathbf{S}_{\rm F}\mathbf{v}^k\|_2\!-\!\sqrt{\beta}\frac{\rho}{\ell N}\|\mathbf{S}_{\rm
               D}\mathbf{v}^k\|_2}{2(\beta\|\mathbf{S}_{\rm F}\mathbf{v}^k\|_2\!+\!\sqrt{\beta}\|\mathbf{S}_{\rm D}\mathbf{v}^k\|_2)}\!\bigg\}$. \\
               \hspace{0.3cm} 1.3 \ Compute \\
                \hspace{0.5cm}$\mathbf{c}^{k+1} = \big(\mathbf{S}_{\rm D}+\frac{\rho}{\ell N}\mathbf{I} + 2\mu^{k*}(\mathbf{S}_{\rm F}-\beta\mathbf{S}_{\rm D})\big)^{\dagger}\mathbf{v}^k$. \\
               \hspace{0.1cm} S.2 \ Solve the subproblem \eqref{Relax x_update}. \\
               \hspace{0.3cm} 2.1 \ Compute $\displaystyle\mathbf{b}^k=\mathbf{w}^k-\frac{\mathbf{y}_2^k}{\rho}$. \\
               \hspace{0.3cm} 2.2 \ Compute $\mathbf{z}^{k+1}$ through the binary section\\
                \hspace{0.5cm}searching procedure in Table \ref{bisection}. \\
               \hspace{0.3cm} 2.3 \ Compute $t^{k+1}=\max\{0, {\rm Re}\big(\mathbf{z}^{k+1 H}\mathbf{b}^k\big)\}$. \\
               \hspace{0.3cm} 2.4 \ Compute $\mathbf{x}^{k+1} = t^{k+1}\mathbf{z}^{k+1}$. \\
               \hspace{0.1cm} S.3 \ Solve the subproblem \eqref{Relax u_update}.\\
               \hspace{0.3cm} 3.1 \ Compute\\
                \hspace{0.5cm}$\mathbf{u}^{k+1} = \displaystyle \frac{\mathbf{y}_1^k+\tilde{\rho}\mathbf{x}^{k+1}+(\rho+\tilde{\rho}){\rm IFFT}_{\ell}(\mathbf{c}^{k+1})}{2\tilde{\rho}+\rho}.$\\
               \hspace{0.3cm} 3.2 \ Compute\\
                \hspace{0.5cm}$\mathbf{w}^{k+1} = \displaystyle \frac{\mathbf{y}_2^k+(\tilde{\rho}+\rho)\mathbf{x}^{k+1}+\tilde{\rho}{\rm IFFT}_{\ell}(\mathbf{c}^{k+1})}{2\tilde{\rho}+\rho}$. \\
               \hspace{0.1cm} S.4 \ Update Lagrangian multipliers. \\
               \hspace{0.3cm} 4.1 \ Compute ${{\mathbf{y}}_1^{k+1}}=\mathbf{y}_1^k + \rho({\rm IFFT}_{\ell}(\mathbf{c}^{k+1})-\mathbf{u}^{k+1})$. \\
               \hspace{0.3cm} 4.2 \ Compute ${\mathbf{y}}_2^{k+1}=\mathbf{y}_2^k + \rho(\mathbf{x}^{k+1}-\mathbf{w}^{k+1})$. \\
               \textbf{\ Until} some preset termination conditions are satisfied. \\
               \hspace{0.9cm} Then, let $\mathbf{x}^{k+1}$ be the output.
     \end{Bflushleft}
     }
     \end{spacing}
     \end{center}
     \caption{ADMM-Relax algorithm for the model \eqref{ori model}.}
     \label{ADMM-Relax}
\end{figure}

\subsection{Performance Analysis}

\subsubsection{Convergence issue}
  We have Theorem \ref{Relax convergence} to show the convergence properties of the proposed ADMM-Relax algorithm \eqref{Relax ADMM iteration}. Its proof is shown in Appendix D.

\begin{theorem}\label{Relax convergence}
    Let $\{\mathbf{c}^k, \mathbf{x}^k$, $\mathbf{u}^k$, $\mathbf{w}^k$, $\mathbf{y}_1^k$, $\mathbf{y}_2^k, k=1,2,\dotsb\}$ be the sequence generated by the proposed ADMM-Relax algorithm \eqref{Relax ADMM iteration} as shown in Figure \ref{ADMM-Relax}. If $\rho> 2\tilde{\rho}$, then
    \begin{itemize}
    \item sequence $\{\mathbf{c}^k, \mathbf{x}^k$, $\mathbf{u}^k$, $\mathbf{w}^k$, $\mathbf{y}_1^k$, $\mathbf{y}_2^k\}$ is convergent, i.e.,
  \begin{equation}\label{Relax converged results}
   \begin{split}
   &\underset{k\rightarrow+\infty}\lim\mathbf{c}^k=\mathbf{c}^*, \ \ \ \underset{k\rightarrow+\infty}\lim\mathbf{x}^k=\mathbf{x}^*, \\
   &\underset{k\rightarrow+\infty}\lim\mathbf{u}^k=\mathbf{u}^*, \ \ \ \underset{k\rightarrow+\infty}\lim\mathbf{w}^k=\mathbf{w}^*, \\
   &\underset{k\rightarrow+\infty}\lim\mathbf{y}_1^k=\mathbf{y}_1^*, \ \ \ \underset{k\rightarrow+\infty}\lim\mathbf{y}_2^k=\mathbf{y}_2^*,
   \end{split}
  \end{equation}
  and $\mathbf{Ac}^*=\mathbf{u}^*, \ \mathbf{x}^*=\mathbf{w}^*, \ \mathbf{y}_1^*=-\mathbf{y}_2^*$.
  \item $(\mathbf{c}^*,\mathbf{x}^*,\mathbf{u}^*,\mathbf{w}^*)$ is some KKT point of the model \eqref{relax model}.
  \item  If $(\mathbf{c}^1, \mathbf{x}^1)$ lies in the feasible region of the model \eqref{ori model}, $(\mathbf{c}^*, \mathbf{x}^*)$ approaches some KKT point of the model \eqref{ori model} as $\tilde{\rho}$ increases.
  \end{itemize}
\end{theorem}

{\it Remarks:}
Theorem \ref{Relax convergence} shows that the proposed ADMM-Relax algorithm is theoretically guaranteed to be convergent. Especially, its third part indicates that if $(\mathbf{c}^1, \mathbf{x}^1)$ lies in the feasible region of the original model \eqref{ori model}, $(\mathbf{c}^*, \mathbf{x}^*)$ approaches some KKT point of the model \eqref{ori model} as $\tilde{\rho}$ increases.
Moreover, in the simulation section, we also show that the residual error, $\lVert\mathbf{Ac}^*-\mathbf{x}^*\rVert_2^2$, of the KKT equations decreases as $\tilde{\rho}$ increases.
Furthermore, the key to prove Theorem \ref{Relax convergence} is to  exploit the unconstrained auxiliary variables $\mathbf{u}$ and $\mathbf{w}$, the augmented Lagrangian function can be guaranteed {\it sufficient descent} in every iteration. However, in ADMM-Direct, the corresponding augmented Lagrangian function cannot be proved to have this kind of property. The detailed proof of Theorem \ref{Relax convergence} can be found in Appendix B.

Moreover, the relaxation does not cause larger PAPR values than $\alpha$ since we let the optimized $\mathbf{x}$ be the final output.
In the above theorem, we mention that, to guarantee the convergence of ADMM-Relax, $\rho$ and $\tilde{\rho}$ should satisfy $\rho>2\tilde{\rho}>0$. Besides that, we should note that there is no theoretical results to help us to set their values. However, it can be seen that the relaxed optimization problem \eqref{relax model} could be ill-conditional if $\tilde{\rho}$ is too large. In the classical augmented Lagrangian multiplier method, there is similar problem on how to set penalty factor. In practice, a general way, but heuristic, to choose proper $\tilde{\rho}$ is to perform simulations when different $\tilde{\rho}$ are set and then select the value corresponding to the best simulation result. In the simulation section, we choose $\tilde{\rho}=100$ and $\rho=300$, which leads to pretty good optimization results.

\subsubsection{Iteration complexity}
We use the residual error which is defined as $\|\mathbf{u}^{k+1}-\mathbf{u}^k\|_2^2 +\|\mathbf{w}^{k+1}-\mathbf{w}^k\|_2^2$ to measure the convergence progress of the ADMM-Relax algorithm since it converges to zero as $k\rightarrow+\infty$. Then, we have Theorem \ref{iteration complexity} about its convergence progress. The detailed proof is shown in Appendix E.


\begin{theorem}\label{iteration complexity}
  Let $r$ be the minimum iteration index such that $\|\mathbf{u}^{k+1}-\mathbf{u}^k\|_2^2 +\|\mathbf{w}^{k+1}-\mathbf{w}^k\|_2^2\leq\epsilon$, where $\epsilon$ is the desired precise parameter for the solution. Then, we have the following iteration complexity result
  \vspace{-3pt}
 \[
   \begin{split}
     &r \leq \frac{1}{C\epsilon}\bigg({L}_{\rho}(\mathbf{c}^{1},\mathbf{x}^{1}, \mathbf{u}^{1},\mathbf{w}^{1}, \mathbf{y}_1^{1}, \mathbf{y}_2^{1}) \\ &\hspace{1.5cm} -\big(\frac{1}{2}\|\mathbf{S}_{\rm D}(\mathbf{c}^*-\mathbf{c}_{\rm o})\|_2^2 + \frac{\tilde{\rho}}{2}\|\mathbf{u}^*-\mathbf{w}^*\|_2^2\big)\bigg),
   \end{split}
 \]
 \vspace{-3pt}
 where $\rho>2\tilde\rho$ and the constant $C$ is the minimum eigenvalue of the following positive definite matrix
 \[
  \begin{bmatrix} \frac{\tilde\rho+\rho}{2}-\frac{2\tilde{\rho}^2}{\rho} & \ \frac{2\tilde{\rho}^2}{\rho}-\frac{\tilde{\rho}}{2}\\ \frac{2\tilde{\rho}^2}{\rho}-\frac{\tilde{\rho}}{2} & \frac{\tilde\rho+\rho}{2}-\frac{2\tilde{\rho}^2}{\rho}\\
\end{bmatrix}.
\]
\end{theorem}

\subsubsection{Computational cost}
   In Figure \ref{ADMM-Relax}, we still use operators ${\rm IFFT}_{\ell}(\cdot)$ and ${\rm FFT}_{\ell}(\cdot)$ to take the place of $\mathbf{A}$ and $\mathbf{A}^H$ respectively.
  Similar to the computational complexity analysis for ADMM-Direct, we can conclude that the computational cost in each ADMM-Relax iteration is roughly $\mathcal{O}(\ell N\log_2\ell N)$.
  Combining this result with Theorem \ref{iteration complexity}, we conclude that the total computational cost to attain an $\epsilon$-optimal solution is $\mathcal{O}(\lfloor d\rfloor\ell N\log_2\ell N)$, where $d=\frac{1}{C\epsilon}\big({L}_{\rho}(\mathbf{c}^{1},\mathbf{x}^{1}, \mathbf{u}^{1},\mathbf{w}^{1}, \mathbf{y}_1^{1}, \mathbf{y}_2^{1})-(\frac{1}{2}\|\mathbf{S}_{\rm D}(\mathbf{c}^*-\mathbf{c}_{\rm o})\|_2^2 + \frac{\tilde{\rho}}{2}\|\mathbf{u}^*-\mathbf{w}^*\|_2^2)\big)$.

\section{Simulation Results}
    In this section, several simulation results are presented to illustrate the performance of the proposed ADMM-Direct and ADMM-Relax.
    We compare the proposed algorithms with the RCF approach \cite{clipping}, Aggarwal SOCP approach \cite{SOCP}, simplified OICF approach \cite{simplified_OICF}, modified SLM approach \cite{modified_SLM} and low-complexity tone injection scheme \cite{TI}.

   Throughout this section, simulation parameters are set as follows: PAPR constraint, for ADMMs, is 4.0dB.
   Free carrier power overhead, for ADMMs, is 0, 0.15, and 0.3. The penalty parameters for ADMM-Direct are $\rho=100$, and for ADMM-Relax they are $\rho=300$ and $\tilde\rho=100$. Consider an OFDM scheme\footnotemark~with 52 data carriers and 12 free carriers and the over-sampling factor $\ell=4$. For all of the bit error ratio (BER) simulations, we calculate $\rm {E}_b$ by
   \footnotetext{The considered OFDM scheme is based on IEEE 802.11a/g Wi-Fi standard. The proposed ADMM-Direct/-Relax algorithms can also be applied to reducing PAPR values of OFDM signals in the 4G/5G cellular systems. But the simulation parameters, such as $\alpha$, $\beta$, $\rho$, and $\tilde{\rho}$, should be re-chosen carefully to achieve desired system performance. Moreover, since the proposed algorithms have much cheaper computational complexity in each iteration than state-of-the-art PAPR reduction approaches, they could be more suitable for large-scale OFDM system.}
\[
{\rm E_b}=\frac{{{\rm \bar{E}_s}}}{\rm M\cdot mod\_style},
\]
where $ {\rm {\bar E}_s}$ represents the averaged energy of the optimal frequency-domain OFDM symbols, $M$ is the number of data carriers, and ``${\rm mod\_style}$'' represents the modulation scheme, which is 2 and 4 corresponding to QPSK and 16-QAM modulation respectively. In the simulations, the number of OFDM symbols are 5000. Moreover, all the simulations are implemented in MATLAB 2017 environment.

\begin{figure}[htbp]
    \begin{minipage}[c]{1.0\linewidth}
    \centering
    \centerline{\includegraphics[width=9.5cm]{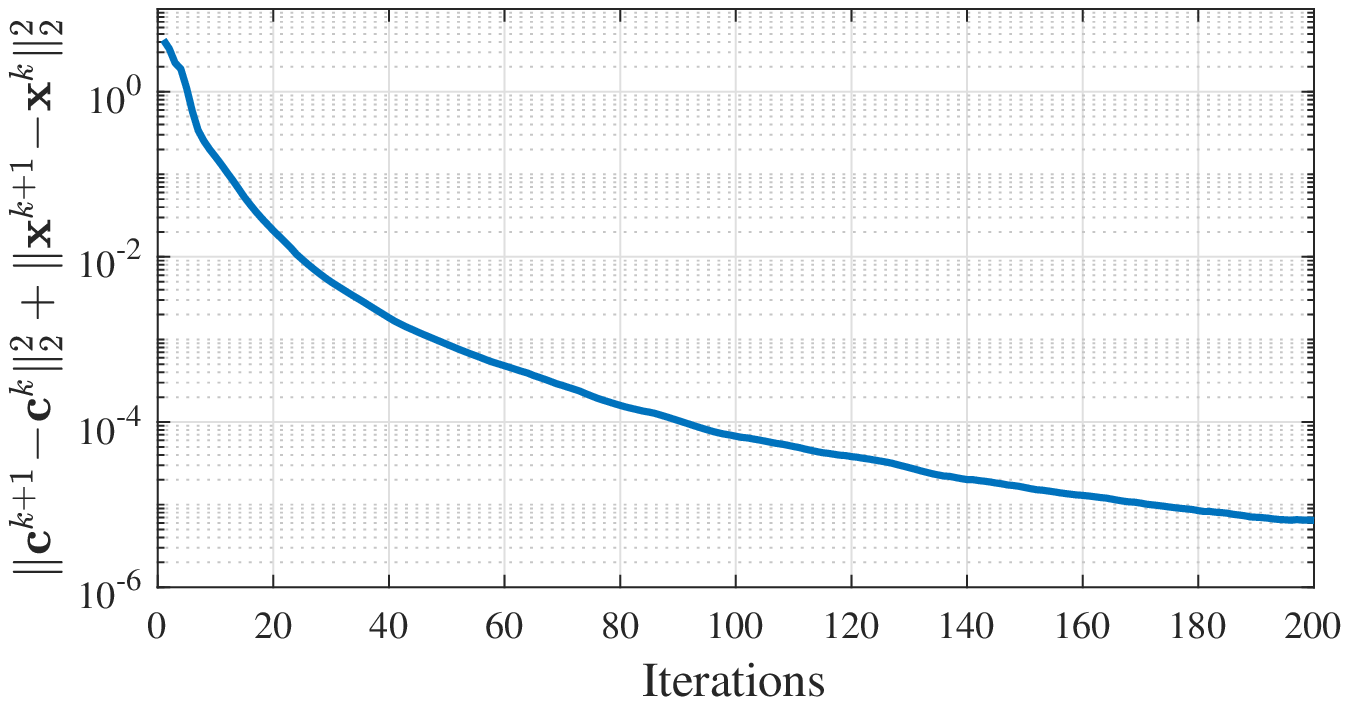}}
    \centerline{(a) ADMM-Direct}\medskip
    \end{minipage}
    \begin{minipage}[c]{1.0\linewidth}
    \centering
    \centerline{\includegraphics[width=9.5cm]{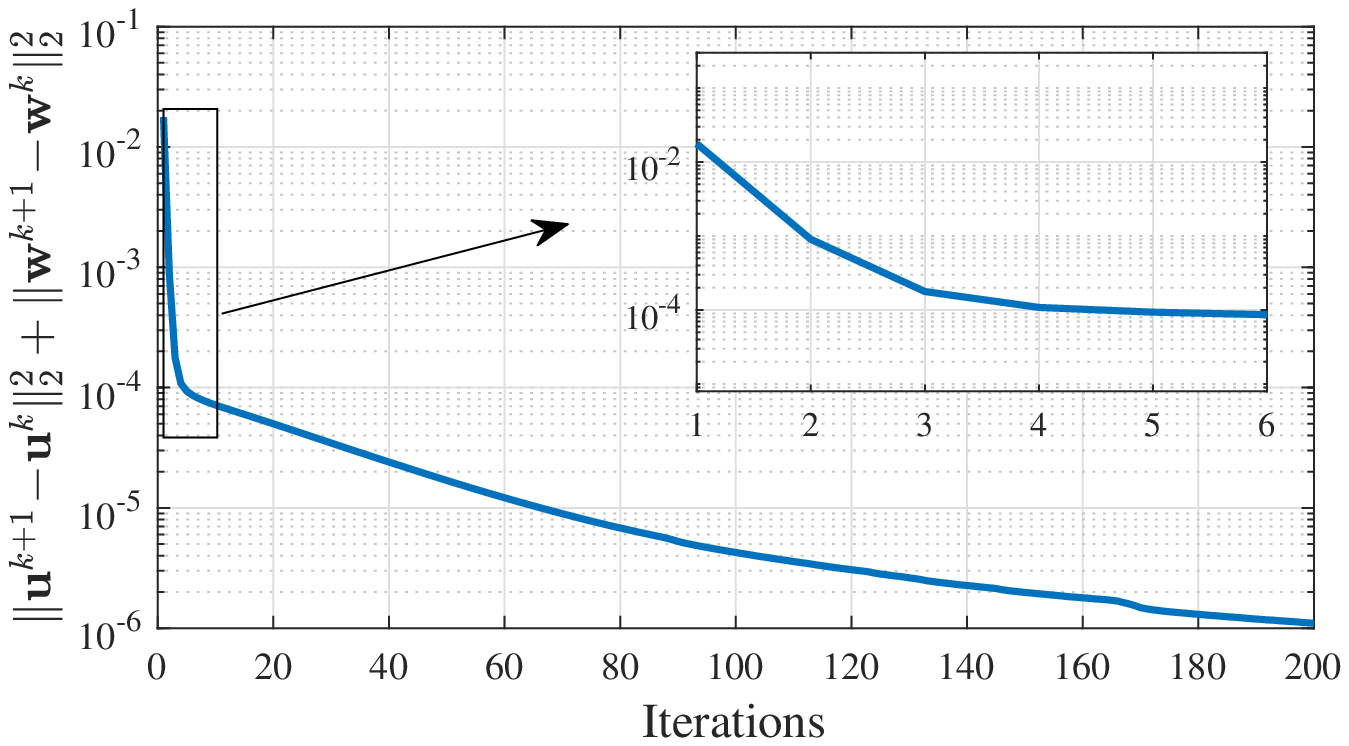}}
    \centerline{(b) ADMM-Relax}\medskip
    \end{minipage}
    \caption{The convergence performance of ADMM-Direct and ADMM-Relax with 16-QAM modulation.}
    \label{Converge}
    \end{figure}

    Figure~\ref{Converge} shows the convergence curves of the proposed ADMM-Direct and ADMM-Relax algorithms. In Figure~\ref{Converge}(a), the residual error of ADMM-Direct is defined as $\|\mathbf{c}^{k+1}-\mathbf{c}^k\|_2^2+\|\mathbf{x}^{k+1}-\mathbf{x}^k\|_2^2$, and in Figure~\ref{Converge}(b), the residual error of ADMM-Relax is defined as $ \|\mathbf{u}^{k+1}-\mathbf{u}^k\|_2^2+\|\mathbf{w}^{k+1}-\mathbf{w}^k\|_2^2$.
    From the curves, we can see that both of the ADMMs can converge after a few iterations.
    Here, we should note that we do not give the exact proof of the convergence for the ADMM-Direct algorithm.
    However, we observe that ADMM-Direct can converge from Figure~\ref{Converge}(a).
    From Figure~\ref{Converge}(b), we can see that the residual error decreases quickly in the first several iterations and after 5
iterations, the convergence curve is relatively flat.

    \begin{figure}[htbp]
     \centering
     \includegraphics[width=9.5cm]{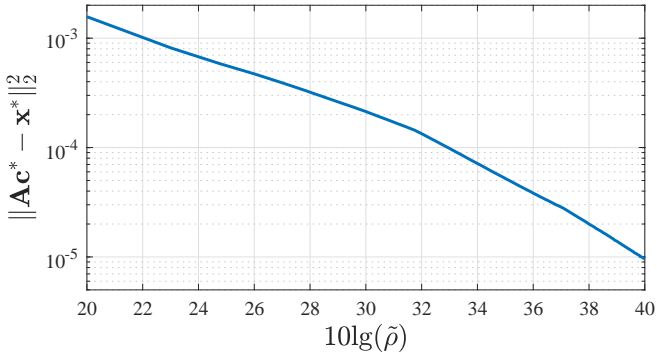}
     \caption{The relationship between $\|\mathbf{Ac}^*-\mathbf{x}^*\|^2_2$ and $\tilde{\rho}$.}
     \label{residual_error}
     \end{figure}

To further illustrate the third part of Theorem \ref{Relax convergence}, which indicates that if $(\mathbf{c}^1, \mathbf{x}^1)$ lies in the feasible region of the model \eqref{ori model}, $(\mathbf{c}^*, \mathbf{x}^*)$ approaches some KKT point of the model \eqref{ori model} as $\tilde{\rho}$ increases, we plot the curve of the relationship between $\|\mathbf{Ac}^*-\mathbf{x}^*\|^2_2$ and $\tilde{\rho}$ in Figure~\ref{residual_error}.
In this figure, $(\mathbf{c}^*, \mathbf{x}^*)$ is the optimal solution of the model \eqref{relax model}.
Moreover, we use $\|\mathbf{Ac}^*-\mathbf{x}^*\|^2_2$ to measure whether $(\mathbf{c}^*, \mathbf{x}^*)$ approaches some KKT point of the model \eqref{ori model}, because if $\|\mathbf{Ac}^*-\mathbf{x}^*\|^2_2$ approaches zero, $(\mathbf{c}^*, \mathbf{x}^*)$ approaches some KKT point of the model \eqref{ori model} as Appendix D shows.
From the curve, we can see that, $\|\mathbf{Ac}^*-\mathbf{x}^*\|^2_2$ approaches zero as $\tilde{\rho}$ increases as we expected.
That is, $(\mathbf{c}^*, \mathbf{x}^*)$ approaches some KKT point of \eqref{ori model}.

\begin{table}[htbp]

\renewcommand \arraystretch{1.0}
\centering
\caption{Comparison of EVMs (dB) at different $\beta$ (Modulation: 16-QAM; PAPR constraint: 4dB)}
\begin{tabular}{|c||c|c|}
\hline \hline
       \multirow{2}{*}{$\beta$}  &\multicolumn{2}{c|}{EVM} \\
        \cline{2-3}
        &ADMM-Direct &ADMM-Relax \\
        \hline
       0    &  -16.58dB         & -16.36dB \\ \hline
       0.15 &  -27.33dB         & -27.51dB \\ \hline
       0.3  &  -32.96dB         & -32.89dB \\ \hline \hline

\end{tabular}
\end{table}

  In Table II, we show that the impact of the different values of $\beta$ on signal distortion introduced to data carriers of OFDM signals. Here, we use averaged error vector magnitude (EVM) to evaluate distortion, which is defined by
    \[
  \overline{\rm EVM} = \sqrt{\frac{1}{K}\displaystyle\sum_{i=1}^{K} \frac{\|\mathbf{S}_{\rm D}(\mathbf{c}-\mathbf{c}_{\rm o})\|^2_2}{\|\mathbf{c}_{\rm o}\|^2_2}},
   \]
  where $K$ is set 5000 in the simulations.
    From Table II, we can see clearly that the signal distortion decreases almost 10dB when we increase $\beta$ from 0 to 0.15. However, when we further increase $\beta$ from 0.15 to 0.3, the corresponding EVM values, i.e., introduced distortions, only decrease about 5.5dB. Checking the optimized free carriers, we find that the power overheads of some OFDM symbols in the latter are less than the pre-set upper-bound 0.3. It means that the constraint 3(c) becomes inactive for these kinds of OFDM symbols and their optimizers locate inside the defined feasible region. The fact indicates that increasing $\beta$ can decrease averaged signal distortion of the optimized OFDM symbols efficiently. However, for a larger $\beta$, its influence becomes weaker.

\begin{figure}[htbp]
    \begin{minipage}[c]{1.0\linewidth}
    \centering
    \centerline{\includegraphics[width=9.5cm]{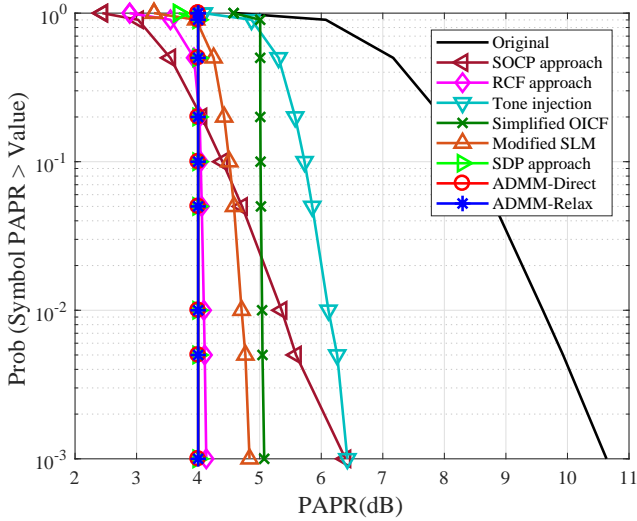}}
    \centerline{(a) PAPR reduction performance}\medskip
    \label{16-QAM_PAPR}
    \end{minipage}
    \begin{minipage}[c]{1.0\linewidth}
    \centering
    \centerline{\includegraphics[width=9.5cm]{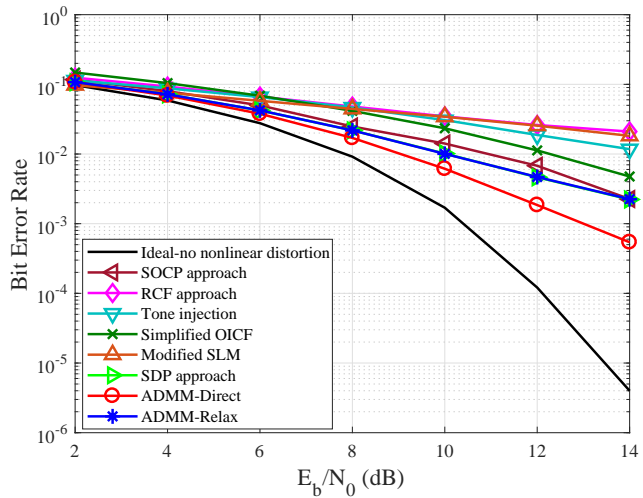}}
    \centerline{(b) BER performance (after SSPA)}\medskip
    \label{16-QAM_Pe}
    \end{minipage}
    \begin{minipage}[c]{1.0\linewidth}
    \centering
    \centerline{\includegraphics[width=9.5cm]{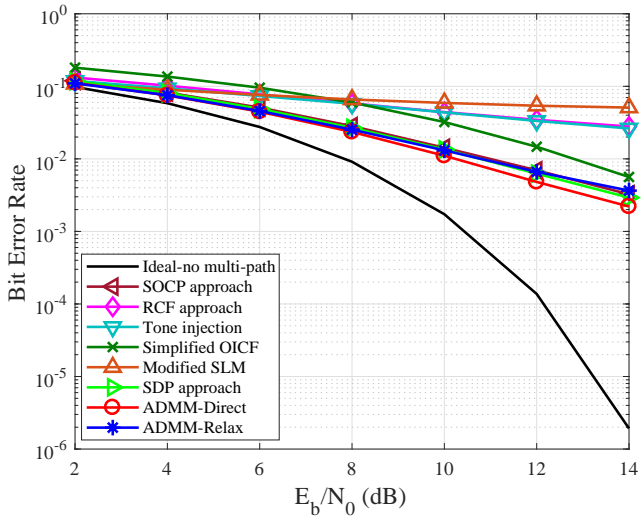}}
    \centerline{(b) BER performance (through multi-path channel)}\medskip
    \label{16-QAM_Pe_multipath}
    \end{minipage}
    \caption{The performance of various systems with 16-QAM modulation(5 iterations for ADMMs, 10 iterations for RCF).}
    \label{16-QAM}
    \end{figure}

\begin{figure}[htbp]
    \begin{minipage}[c]{1.0\linewidth}
    \centering
    \centerline{\includegraphics[width=9.5cm]{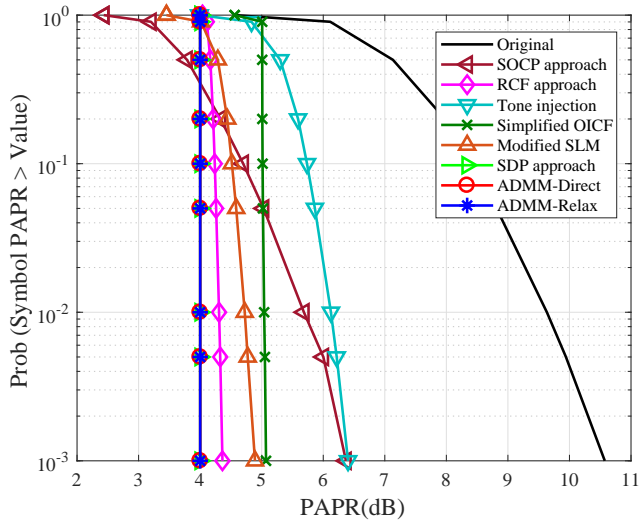}}
    \centerline{(a) PAPR reduction performance}\medskip
    \label{QPSK_PAPR}
    \end{minipage}
    \begin{minipage}[c]{1.0\linewidth}
    \centering
    \centerline{\includegraphics[width=9.5cm]{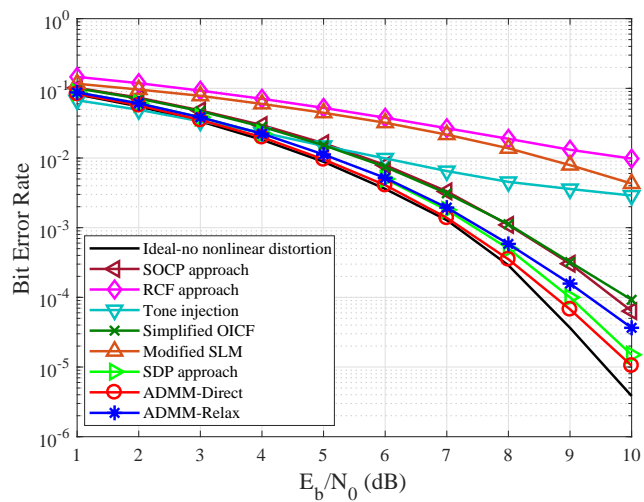}}
    \centerline{(b) BER performance (after SSPA)}\medskip
    \label{QPSK_Pe}
    \end{minipage}
    \begin{minipage}[c]{1.0\linewidth}
    \centering
    \centerline{\includegraphics[width=9.5cm]{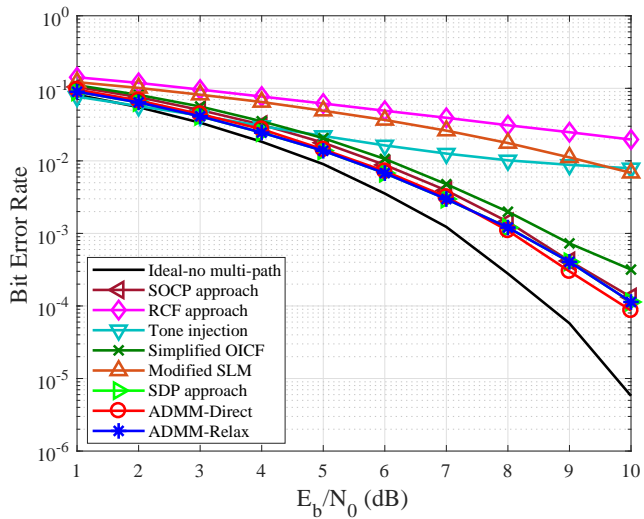}}
    \centerline{(b) BER performance (through multi-path channel)}\medskip
    \label{QPSK_Pe_multipath}
    \end{minipage}
    \caption{The performance of various systems with QPSK modulation(5 iterations for ADMMs, 10 iterations for RCF).}
    \label{QPSK}
    \end{figure}

    Figure~\ref{16-QAM} and Figure~\ref{QPSK} plot PAPR complementary cumulative distribution functions (CCDFs), bit error rate (BER) performance of the original and processed OFDM signals after the solid state PA (SSPA) with smoothing factor 3 (modeled in \cite{Tutorial}) and through AWGN channel, and BER performance of OFDM signals after SSPA and through multi-path channel. Data carrier modulations are assumed to be 16-QAM or QPSK.
    CCDF denotes the probability that the PAPR of the OFDM symbols exceeds some given threshold $T$, i.e.,
      ${\rm CCDF}({T}) = {\rm Prob}({\rm PAPR}>{T})$.
    The simulation results of ADMMs are obtained through 5 iterations.

\begin{figure}[htbp]
     \centering
     \includegraphics[width=9.5cm]{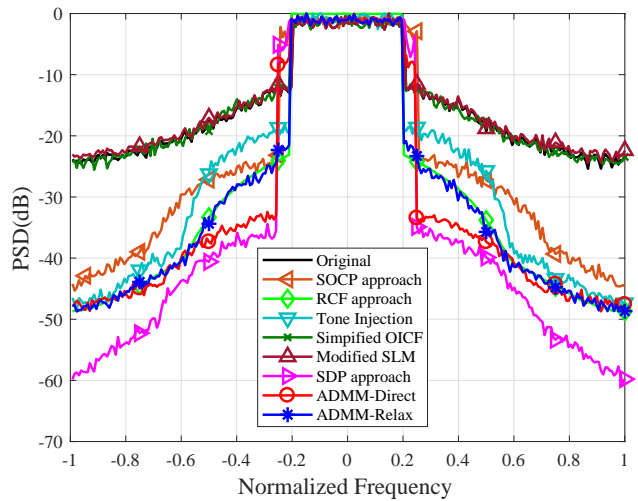}
     \caption{Out-of-band radiation performance with 16-QAM modulation.}
     \label{PSD}
\end{figure}

    In Figure~\ref{16-QAM}(a), the word ``Original'' means no distortion is introduced. From the figure, we can see that the proposed ADMMs, RCF, OICF, and SDP, have cut-off CCDF curves. However, the CCDF curves of ADMMs and SDP locate on the left side of the others'. It means that the former three approaches have better PAPR reduction performance. Moreover, CCDF curves of SOCP, modified SLM and tone injection are slow-down. It means that some of their optimized OFDM symbols still have larger PAPR values. In practice, these kinds of signals would suffer from severe nonlinear distortion of the PA, which can worse BER performance of the OFDM signals. Figure~\ref{16-QAM}(b) shows the BER curves of the optimized OFDM symbols after SSPA. The input power back-off of the working-point away from saturation region is set as 4.1dB. From it, we can see that BER curves of ADMMs and SDP are closest to the ideal's. Here, the word ``ideal'' means no distortion is introduced in the OFDM symbols.
    Figure~\ref{QPSK} plots the PAPR-reduction performance and BER performance of the original OFDM signals and the processed OFDM signals with QPSK data carrier modulation.
    Similar to the performance with 16-QAM modulation, ADMM-Direct and ADMM-Relax still have cut-off PAPR reduction performance. Meanwhile, from
    Figure~\ref{QPSK}(b), we can see that the ADMM-Direct and ADMM-Relax have pretty good BER performance. Here, the input power back-off  is also set as 4.1dB.
    Furthermore, Figure~\ref{16-QAM}(c) and \ref{QPSK}(c) plot the BER curves of the optimized OFDM symbols after SSPA and multi-path channel. Here, four paths are considered and in each path delay/fading parameters are $(0,1)$ (direct path), $(190, 0.2)$, $(300, 0.07)$, and $(400, 0.05)$ respectively. From the figures, we can see that, in comparison with AWGN channel case, BER performance of all PAPR reduction methods becomes worse when multi-path effects are considered. However, ADMM-Direct/-Relax approaches are still better than state-of-the-art approaches.
    At last, from Figure \ref{16-QAM} and \ref{QPSK}, we see that SDP approach has similar PAPR reduction performance and BER performance to our proposed ADMM approaches. However, we should note that the computational complexity of SDP is prohibitive in practice, which is roughly $\mathcal{O}(\ell^3N^3)$ in each iteration. In comparison, the computational complexity of our ADMMs is roughly $\mathcal{O}(\ell N \log_2(\ell N))$ in each iteration, which is much cheaper than SDP.
   The out-of-band radiation performance is shown in Figure~\ref{PSD}.
   On the one hand, we can see from the curves that SDP, ADMM-Direct/-Relax and RCF have lower out-of-band emission (OOBE). On the other hand, we can also observe that tone injection and modified SLM have higher OOBE.
   The reason that OOBE performance of the proposed ADMM-Direct/-Relax approaches and SDP appraoch is better because all of them have cut-off CCDF curves, which means that their processed OFDM symbols have quasi-constant PAPR values. It is well known that OOBE is mainly caused by nonlinear distortion of the OFDM signals. In the simulations, since the input power back-off of the working-point is set as 4.1dB and the PAPR of OFDM symbols optimized by ADMM-Direct/-Relax are almost 4dB (quasi-constant), it means that most of the signals are amplified in the linear region of the PA. So, there is only very small nonlinear distortion (caused by nonlinear PA) introduced into the OFDM symbols. Accordingly, it is reasonable that their OOBEs are low.

\section{Conclusion}

     This paper proposes two low-complexity iteration optimization methods named  ADMM-Direct and ADMM-Relax, to reduce PAPR values of the OFDM signals. Both of their computational complexities in each iteration are similar to the classical RCF method and the desired OFDM symbols can be obtained just after a few iterations; meanwhile, they have quasi-constant PAPR values and optimized signal distortion. Moreover, the resulting solution of ADMM-Direct is guaranteed to be some KKT point of the established model when the iteration algorithm converges.
     We also prove that ADMM-Relax is convergent and can approach arbitrarily close to some KKT point of the model if proper algorithm parameters are chosen.
     In comparison with existing algorithms, the proposed ADMM algorithms outperform the existing approaches not only in the simulation results but also in strong theoretically guaranteed performance.
     In the end, we should mention that high PAPR problem is still an issue in the systems of multiple-input multiple-output (MIMO) and non-orthogonal multiple access (NOMA) \cite{NOMA} when multi-carriers techniques are applied. Designing low-complexity, but theoretically guaranteed, ADMM-like optimization algorithm could be an interesting research topic in the future.

\appendices

\section{Proof of equivalence between models \eqref{DIR x model z t} and \eqref{DIR x model z convex}}

The equivalence between the models \eqref{DIR x model z t} and \eqref{DIR x model z convex} is in the sense that the global optimal solution of the latter model is always attained when ``='' holds in \eqref{DIR x model convex z_c}. We prove this fact by contradiction.
To be clear, we suppose some feasible point $\tilde{\mathbf{z}}$ satisfies $\|\tilde{\mathbf{z}}\|_2^2<1$ and the constraint \eqref{DIR x model convex z_b}.
Notice that the constant modulus $\tilde{\mathbf{z}}$ is in the feasible region since $\alpha$ is greater than 1.
 Therefore, we can always find a vector $\bigtriangleup \tilde{\mathbf{z}}$, whose phase vector is equal to that of vector $\mathbf{b}^k$, which can let $\tilde{\mathbf{z}}+\bigtriangleup\tilde{\mathbf{z}}$ satisfy the constraints \eqref{DIR x model convex z_b} and \eqref{DIR x model convex z_c}.
It is obvious that the new vector corresponds to a larger objective value, which means that $\tilde{\mathbf{z}}$ is not the global optimal solution of the model \eqref{DIR x model z convex}.
So, we can conclude that the maximizer in model \eqref{DIR x model z convex} should always satisfy \eqref{DIR x model convex z_c} when ``='' holds.

\section{Proof of Theorem \ref{convergence ADMM DIR}}
   To be clear, we let $L(\mathbf{c},\mathbf{x}, \gamma, \mu,\mathbf{y})$, $L^{\mathbf{c}}(\mathbf{c}, \mathbf{y}, \mu)$, and $L^{\mathbf{x}}(\mathbf{x}, \mathbf{y}, \gamma)$ be the Lagrangian functions of the model \eqref{ori model}, \eqref{DIR c_update}, and \eqref{DIR x_update} respectively.
   $\mu$ and $\gamma$ are the Lagrangian multipliers corresponding to the constraints \eqref{ori model_c} and \eqref{ori model_b} respectively.
   We also let $\mu^*$ and $\gamma^*$ be the corresponding optimal Lagrangian multipliers when ADMM-Direct algorithm is convergent.
   To show $(\mathbf{c}^*,\mathbf{x}^*)$ is a KKT point, we should prove that it, combining ($\mathbf{y}^*$,  $\mu^*$, $\gamma^*$ ), should satisfy the conditions of the primal feasibility \eqref{ori pri feasibility}, the dual feasibility \eqref{ori dual feasibility} and the complementary slackness \eqref{ori complementary slackness}, and is also a stationary point of the Lagrangian function  $L(\mathbf{c},\mathbf{x}, \gamma, \mu, \mathbf{y})$,
   i.e.,
   \begin{subequations}\label{ORI KKT}
    \begin{align}
      & \mathbf{c^*\in\mathcal{C}}, \ \ \mathbf{x^*\in\mathcal{X}}, \label{ori pri feasibility}\\
      & \mu^*\geq0, \label{ori dual feasibility} \\
      & \mu^*(\|\mathbf{S}_{\rm F}\mathbf{c}^*\|_2^2-\beta\|\mathbf{S}_{\rm D}\mathbf{c}^*\|_2^2)=0, \label{ori complementary slackness} \\
      &\nabla_\mathbf{c} {L}(\mathbf{c}^*,\mathbf{x}^*,\gamma^*,\mu^*,\mathbf{y}^*)=0, \label{ori stationary 1}\\
      &\nabla_\mathbf{x} L(\mathbf{c}^*,\mathbf{x}^*,\gamma^*,\mu^*,\mathbf{y}^*)=0, \label{ori stationary 2}
    \end{align}
  \end{subequations}
  where $\mathcal{X}$ and $\mathcal{C}$ denote the constraints \eqref{ori model_b} and \eqref{ori model_c} respectively.

 Since in every ADMM-Direct iteration, $\mathbf{c}^{k+1}$ and $\mathbf{x}^{k+1}$ are located in the feasible region, we can see that the primal feasibility condition \eqref{ori pri feasibility} is satisfied.
 Since $\mu^{k*}$ is guaranteed to be greater than zero (see \eqref{DIR mu_c}) in every  iteration, it means that $ \mu^{*}$ is also greater than zero.
 Moreover, checking \eqref{DIR mu_c} again, we see that the value of $\mu^{k*},\ \forall k,$ is nonzero or zero corresponding to the constraint \eqref{ori model_c}, which is active or inactive respectively. It means that $\mu^*$ satisfies the condition of complementary slackness.

 Now, let us consider \eqref{ori stationary 1} and \eqref{ori stationary 2}. Since $\mathbf{c}^{k+1}$ and ${\mathbf{x}^{k+1}}$ are the minimizers of the problems \eqref{DIR c_update} and \eqref{DIR x_update} in the $k$th iteration respectively, they should satisfy
\begin{equation}\label{ori KKT cx}
  \begin{split}
  &\nabla_\mathbf{c} {L}^{\mathbf{c}}(\mathbf{c}^{k+1},\mathbf{y}^k,\mu^{k*})+\rho\mathbf{A}^H(\mathbf{Ac}^{k+1}-\mathbf{x}^k) = 0, \\
  &\nabla_\mathbf{x} {L}^{\mathbf{x}}(\mathbf{x}^{k+1},\mathbf{y}^k,\gamma^{k*})-\rho(\mathbf{Ac}^{k+1}-\mathbf{x}^{k+1}) = 0.
  \end{split}
\end{equation}
Since $\underset{k\rightarrow+\infty}\lim (\mathbf{c}^k, \mathbf{x}^k, \mathbf{y}^k) = (\mathbf{c}^*, \mathbf{x}^*, \mathbf{y}^*)$ and $\mathbf{y}^{k+1}=\mathbf{y}^k+\rho(\mathbf{Ac}^{k+1}-\mathbf{x}^{k+1})$, we can drop the second terms in \eqref{ori KKT cx} as $k\rightarrow+\infty$ and obtain
\begin{equation}\label{ori gradient zero}
  \begin{split}
  &\nabla_\mathbf{c} {L}^{\mathbf{c}}(\mathbf{c}^*,\mathbf{y}^*,\mu^{*}) = 0, \ \
  \nabla_\mathbf{x} L^{\mathbf{x}}(\mathbf{x}^*,\mathbf{y}^*,\gamma^{*}) = 0.
  \end{split}
\end{equation}
Since there are
$\nabla_\mathbf{c} {L}^{\mathbf{c}}(\mathbf{c}^*,\mathbf{y}^*,\mu^*)=\nabla_\mathbf{c} {L}(\mathbf{c}^*,\mathbf{x}^*,\gamma^*,\mu^*,\mathbf{y}^*)$ and $\nabla_\mathbf{x} {L}^{\mathbf{x}}(\mathbf{x}^*,\mathbf{y}^*,\gamma^*)=\nabla_\mathbf{x} {L}(\mathbf{c}^*,\mathbf{x}^*,\gamma^*,\mu^*,\mathbf{y}^*)$,
we can see that $(\mathbf{c}^*, \mathbf{x}^*, \mathbf{y}^*)$ should satisfy \eqref{ori stationary 1} and \eqref{ori stationary 2}, i.e., it is a stationary point of the Lagrangian function $L(\mathbf{c},\mathbf{x}, \gamma, \mu,\mathbf{y})$. This concludes the proof of Theorem \ref{convergence ADMM DIR}.

\section{Solving the optimization subproblems \eqref{Relax c_update}-\eqref{Relax u_update}}

\subsection{Solving the Subproblem \eqref{Relax c_update}}
 Based on the augmented Lagrangian function \eqref{Relax ALG}, the problem \eqref{Relax c_update} is equivalent to
\begin{subequations}\label{Relax c model}
  \begin{align}
    &\underset{\displaystyle\mathbf{c}\in\mathbb{C}^N}{\rm min}\hspace{0.5cm} \frac{1}{2}\|\mathbf{S}_{\rm D}(\mathbf{c}-\mathbf{c}_{\rm o})\|_2^2 + \frac{\rho}{2}\|\mathbf{Ac}-\mathbf{u}^k + \frac{\mathbf{y}_1^k}{\rho}\|_2^2,             \label{Relax c model_a} \\
    &{\rm subject\ to} \hspace{0.265cm}\|\mathbf{S}_{\rm F}\mathbf{c}\|_2^2-\beta\|\mathbf{S}_{\rm D}\mathbf{c}\|_2^2\leq 0.   \label{Relax c model_b}
  \end{align}
\end{subequations}
Its Lagrangian function can be written as
\begin{equation}\label{Relax lagrangian c}
 \begin{split}
  L(\mathbf{c},\mu^k) &= \frac{1}{2}\|\mathbf{S}_{\rm D}(\mathbf{c}-\mathbf{c}_{\rm o})\|_2^2 + \frac{\rho}{2}\|\mathbf{Ac}-\mathbf{u}^k + \frac{\mathbf{y}_1^k}{\rho}\|_2^2 \\ &\hspace{0.5cm}+ \mu^k\big(\|\mathbf{S}_{\rm F}\mathbf{c}\|_2^2-\beta\|\mathbf{S}_{\rm D}\mathbf{c}\|_2^2\big),
  \end{split}
\end{equation}
where the Lagrangian multiplier is $\mu^k\geq0$.
Since the global optimal solution $\mathbf{c}^{k+1}$, combining the optimal Lagrangian multiplier $\mu^{k*}$ should satisfy $\displaystyle\nabla_\mathbf{c} L(\mathbf{c}^{k+1}, \mu^{k*})=0$, we can get
\begin{equation}\label{Relax c opt mu}
  \mathbf{c}^{k+1} = \big(\mathbf{S}_{\rm D}+\frac{\rho}{\ell N}\mathbf{I} + 2\mu^{k*}(\mathbf{S}_{\rm F}-\beta\mathbf{S}_{\rm D})\big)^{\dagger}\mathbf{v}^k,
\end{equation}
where $\displaystyle\mathbf{v}^k=\mathbf{c}_{\rm o}+\rho\mathbf{A}^H(\mathbf{u}^k-\frac{\mathbf{y}_1^k}{\rho})$.

Moreover, when the constraint \eqref{Relax c model_b} is inactive, $\mathbf{c}^{k+1}$ is located inside the feasible region $\mathcal{C}$. It means that $\mu^{k*}=0$. Otherwise, if the constraint \eqref{Relax c model_b} is active, it means that $\mathbf{c}^{k+1}$ and the optimal Lagrangian multiplier $\mu^{k*}$ should satisfy the constraint \eqref{Relax c model_b} when ``='' holds. Plugging \eqref{Relax c opt mu} into $\|\mathbf{S}_{\rm F}\mathbf{c}^{k+1}\|_2$ and $\|\mathbf{S}_{\rm D}\mathbf{c}^{k+1}\|_2$, we can obtain
\begin{equation}\label{Relax Sf}
  \begin{split}
    &\|\mathbf{S}_{\rm F}\mathbf{c}^{k+1}\|_2 = \frac{\|\mathbf{S}_{\rm F}\mathbf{v}^k\|_2}{\frac{\rho}{\ell N} + 2\mu^{k*}}, \\
    &\|\mathbf{S}_{\rm D}\mathbf{c}^{k+1}\|_2 = \frac{\|\mathbf{S}_{\rm D}\mathbf{v}^k\|_2}{1+\frac{\rho}{\ell N} - 2\mu^{k*}\beta}.
  \end{split}
\end{equation}
Then, we can solve $\mu^{k*}$ as \eqref{Relax mu_k} when $\|\mathbf{S}_{\rm F}\mathbf{c}^{k+1}\|_2^2=\beta\|\mathbf{S}_{\rm D}\mathbf{c}^{k+1}\|_2^2$.
\begin{equation}\label{Relax mu_k}
  \mu^{k*} = \frac{\big(1+\frac{\rho}{\ell N}\big)\|\mathbf{S}_{\rm F}\mathbf{v}^k\|_2-\sqrt{\beta}\frac{\rho}{\ell N}\|\mathbf{S}_{\rm D}\mathbf{v}^k\|_2}{2(\beta\|\mathbf{S}_{\rm F}\mathbf{v}^k\|_2+\sqrt{\beta}\|\mathbf{S}_{\rm D}\mathbf{v}^k\|_2)}.
\end{equation}
Furthermore, observing \eqref{Relax mu_k}, we see that the computed result for $\mu^{k*}$ could be negative. However, the Lagrangian multiplier theory guarantees that $\mu^{k*}$ should always be nonnegative since it is an inequality constraint \eqref{Relax c model_b}. This contradiction comes from the assumption that the constraint is active. It means that the constraint is inactive, therefore $\mu^{k*}$ should be zero. Based on this fact, we compute $\mu^{k*}$ by
\begin{equation}\label{Relax mu_c}
\mu^{k*}\!=\!\max\!\bigg\{\!0,\!\frac{\big(1\!+\!\frac{\rho}{\ell N}\big)\|\mathbf{S}_{\rm F}\mathbf{v}^k\|_2\!-\!\sqrt{\beta}\frac{\rho}{\ell N}\|\mathbf{S}_{\rm D}\mathbf{v}^k\|_2}{2(\beta\|\mathbf{S}_{\rm F}\mathbf{v}^k\|_2+\sqrt{\beta}\|\mathbf{S}_{\rm D}\mathbf{v}^k\|_2)}\bigg\}.
\end{equation}

\subsection{Solving the Subproblem \eqref{Relax x_update}}
The problem \eqref{Relax x_update} can be equivalent to
\begin{subequations}\label{Relax x model}
  \begin{align}
    &\underset{\displaystyle\mathbf{x}\in\mathbb{C}^{\ell N}}{\rm min}\hspace{0.5cm} \|\mathbf{x}-\mathbf{w}^k + \frac{\mathbf{y}_2^k}{\rho}\|_2^2,             \label{Relax x model_a} \\
    &{\rm subject\ to} \hspace{0.5cm} \frac{\|\mathbf{x}\|_\infty^2}
{\frac{1}{\ell N}\|\mathbf{x}\|_2^2}= \alpha .           \label{Relax x model_b}
  \end{align}
\end{subequations}
Similar to the ADMM-Direct algorithm, we also introduce auxiliary variables $t$ and $\mathbf{z}$ to express $\mathbf{x}$ by
$\mathbf{x}=t\mathbf{z}$,
where $t>0$ and $\|\mathbf{z}\|_2^2=1$. Plugging the auxiliary variables $t$ and $\mathbf{z}$ into the problem \eqref{Relax x model}, we can obtain
\begin{subequations}\label{Relax x model t}
  \begin{align}
    &\underset{\displaystyle\mathbf{z}\in\mathbb{C}^{\ell N}, t>0}{\rm min}\hspace{0.3cm} t^2-2t{\rm Re}(\mathbf{z}^H\mathbf{b}^k),             \label{Relax x model t_a} \\
    &\hspace{0.48cm}{\rm subject\ to} \hspace{0.5cm} |z_i|^2 \leq \frac{\alpha}{\ell N}, \ \ i=1,\dotsb,\ell N,   \label{Relax x model t_b} \\
    &\hspace{2.6cm}  \|\mathbf{z}\|_2^2 = 1,
  \end{align}
\end{subequations}
where $
\displaystyle\mathbf{b}^k=\mathbf{w}^{k}-\frac{\mathbf{y}_2^k}{\rho}.
$

To solve $\mathbf{z}^{k+1}$, we can drop $t$ from the model \eqref{Relax x model t} and formulate the following equivalent convex optimization model.
{
\begin{subequations}\label{Relax x model z t}
  \begin{align}
    &\underset{\displaystyle\mathbf{z}\in\mathbb{C}^{\ell N}}{\rm max}\hspace{0.6cm} {\rm Re}(\mathbf{z}^H\mathbf{b}^k),             \label{Relax x model z t_a} \\
    &{\rm subject\ to} \hspace{0.3cm} |z_i|^2 \leq \frac{\alpha}{\ell N}, \ \ i=1,\dotsb,\ell N,    \label{Relax x model z_b} \\
    &\hspace{1.85cm}  \|\mathbf{z}\|_2^2 \leq 1. \label{Relax x model z t_c}
  \end{align}
\end{subequations}}
By introducing the Lagrangian multiplier $\gamma^k>0$ for the constraint \eqref{Relax x model z t_c}, we can change the model \eqref{Relax x model z t} to
\begin{subequations}\label{Relax x model z sep}
  \begin{align}
    &\underset{\displaystyle{z_i}\in\mathbb{C}, \gamma^k>0}{\rm min}\hspace{0.3cm} \displaystyle\sum_{i=1}^{\ell N}-{\rm Re}(z_i^\dagger b_i^k) + {\gamma}^k\big(\displaystyle\sum_{i=1}^{\ell N}|z_i|^2-1\big),    \label{Relax x model z_spe a} \\
    &\hspace{0.48cm} {\rm subject\ to} \hspace{0.6cm} |z_i| \leq \sqrt{\frac{\alpha}{\ell N}}, \ \ i=1,\dotsb,\ell N.   \label{Relax x model z_spe b}
  \end{align}
\end{subequations}
Since both the objective function \eqref{Relax x model z_spe a} and constraint \eqref{Relax x model z_spe b} are treated separately in the variable $z_i$, solving the model \eqref{Relax x model z sep} is equivalent to solving the following $\ell N$ subproblems, which can be performed in parallel.
\begin{subequations}\label{Relax x model z single}
  \begin{align}
    &\underset{\displaystyle{z_i}\in\mathbb{C}, \gamma^k>0}{\rm min}\hspace{0.3cm} -{\rm Re}(z_i^\dagger b_i^k) + \gamma^k|z_i|^2,    \label{Relax x model z_single a} \\
    & \hspace{0.48cm} {\rm subject\ to} \hspace{0.6cm} |z_i| \leq \sqrt{\frac{\alpha}{\ell N}}.   \label{Relax x model z_single b}
  \end{align}
\end{subequations}
Moreover, since only one constraint is involved in \eqref{Relax x model z single}, its optimal solution can be obtained through \eqref{Relax zi}.

\begin{equation}\label{Relax zi}
  z_i^{k+1} = \begin{cases} \hspace{0.5cm} \displaystyle\frac{b_i^k}{2\gamma^k}, \hspace{1cm} \frac{|b_i^k|}{2\gamma^k}<\sqrt\frac{\alpha}{\ell N}, \\
  \sqrt\frac{\alpha}{\ell N}e^{\displaystyle j\phi(b_i^k)}, \ \ {\rm otherwise},
  \end{cases}
\end{equation}
where $\phi(b_i^k)$ represents the phase of $b_i^k$.
Furthermore, since $\mathbf{z}^{k+1}$ should satisfy the constraint
$\|\mathbf{z}^{k+1}\|_2^2=1$, the optimal Lagrangian multiplier $\gamma^{k*}$ can be determined by the binary section searching procedure as shown in Table I. After that, plugging the obtained $\mathbf{z}^{k+1}$ into the model \eqref{Relax x model z t} and simplifying it as a quadratic problem, we can get
$t^{k+1}= {\rm Re}\big(\mathbf{z}^{k+1 H}\mathbf{b}^k\big)$.
At last, plugging $\mathbf{z}^{k+1}$ and $t^{k+1}$ into $\mathbf{x}=t\mathbf{z}$, we get the optimal solution $\mathbf{x}^{k+1}$ of \eqref{Relax x model}.
\subsection{Solving the Subproblem \eqref{Relax u_update}}
Since the problem \eqref{Relax u_update} is an unconstrained quadratic problem, its optimal solution $(\mathbf{u}^{k+1},\mathbf{w}^{k+1})$ should satisfy
\begin{subequations}
\begin{align}
&\nabla_\mathbf{u} {L}_{\rho}(\mathbf{x}^{k+1},\mathbf{c}^{k+1},\mathbf{u}^{k+1},\mathbf{w}^{k+1},\mathbf{y}_1^k,\mathbf{y}_2^k) = 0,\\
&\nabla_\mathbf{w} {L}_{\rho}(\mathbf{x}^{k+1},\mathbf{c}^{k+1},\mathbf{u}^{k+1},\mathbf{w}^{k+1},\mathbf{y}_1^k,\mathbf{y}_2^k) = 0.
\end{align}
\end{subequations}
That is $(\mathbf{u}^{k+1},\mathbf{w}^{k+1})$ is the solution of
{
\begin{subequations}
\begin{align}
&-\mathbf{y}_1^k+\tilde{\rho}(\mathbf{u}^{k+1}-\mathbf{w}^{k+1})-\rho(\mathbf{Ac}^{k+1}-\mathbf{u}^{k+1})=0,\\
&-\mathbf{y}_2^k-\tilde{\rho}(\mathbf{u}^{k+1}-\mathbf{w}^{k+1})-\rho(\mathbf{x}^{k+1}-\mathbf{w}^{k+1})=0.
\end{align}
\end{subequations}}
Solving these two equations, we can get
\begin{subequations}
\begin{align}
&\mathbf{u}^{k+1} = \displaystyle \frac{\mathbf{y}_1^k+\tilde{\rho}\mathbf{x}^{k+1}+(\rho+\tilde{\rho}){\rm IFFT}_{\ell}(\mathbf{c}^{k+1})}{2\tilde{\rho}+\rho}, \label{Relax uopt}\\
&\mathbf{w}^{k+1} = \displaystyle \frac{\mathbf{y}_2^k+(\tilde{\rho}+\rho)\mathbf{x}^{k+1}+\tilde{\rho}{\rm IFFT}_{\ell}(\mathbf{c}^{k+1})}{2\tilde{\rho}+\rho}. \label{Relax wopt}
\end{align}
\end{subequations}

\section{Proof of Theorem \ref{Relax convergence}}

First, we can see that the ADMM-Relax algorithm \eqref{Relax ADMM iteration} can guarantee the resulting $\mathbf{c}^{k+1}$ and $\mathbf{x}^{k+1}$ to satisfy \eqref{relax model_b} and \eqref{relax model_c} respectively. Thus, without loss of generality we assume that $\mathbf{c}^1$ and $\mathbf{x}^1$ satisfy \eqref{relax model_b} and \eqref{relax model_c} respectively.

Since $\mathbf{c}^{k+1}$ and $\mathbf{x}^{k+1}$ are the minimizers of the problems \eqref{Relax c_update} and \eqref{Relax x_update} respectively, we have
\begin{equation}\label{df_c}
  \begin{split}
     &L_{\rho}(\mathbf{c}^k,\mathbf{x}^k,\mathbf{u}^k,\mathbf{w}^k,\mathbf{y}_1^k,\mathbf{y}_2^k)\\
     &\hspace{2cm} -L_{\rho}(\mathbf{c}^{k+1},\mathbf{x}^k,\mathbf{u}^k,\mathbf{w}^k,\mathbf{y}_1^k,\mathbf{y}_2^k)\geq0,
  \end{split}
\end{equation}
\begin{equation}\label{df_x}
  \begin{split}
     &L_{\rho}(\mathbf{c}^{k+1},\mathbf{x}^k,\mathbf{u}^k,\mathbf{w}^k,\mathbf{y}_1^k,\mathbf{y}_2^k) \\&\hspace{2cm} -L_{\rho}(\mathbf{c}^{k+1},\mathbf{x}^{k+1},\mathbf{u}^k,\mathbf{w}^k,\mathbf{y}_1^k,\mathbf{y}_2^k)\geq0.
  \end{split}
\end{equation}
Moreover, based on the Taylor expansion, we can obtain
\begin{equation}\label{Relax u relation}
  \begin{split}
  &L_{\rho}(\mathbf{c}^{k+1},\mathbf{x}^{k+1},\mathbf{u}^k,\mathbf{w}^k,\mathbf{y}_1^k,\mathbf{y}_2^k)\\
  &\hspace{1cm}
  -L_{\rho}(\mathbf{c}^{k+1},\mathbf{x}^{k+1},\mathbf{u}^{k+1},\mathbf{w}^{k+1},\mathbf{y}_1^k,\mathbf{y}_2^k)\\
  =&\frac{1}{2}\begin{bmatrix} \mathbf{u}^{k}-\mathbf{u}^{k+1}\\ \mathbf{w}^k-\mathbf{w}^{k+1} \end{bmatrix}^H
  \begin{bmatrix} \tilde{\rho}+\rho & \ -\tilde{\rho}\\ -\tilde{\rho} & \tilde{\rho}+\rho\\
\end{bmatrix}\begin{bmatrix} \mathbf{u}^{k}-\mathbf{u}^{k+1}\\ \mathbf{w}^k-\mathbf{w}^{k+1} \end{bmatrix},\\
  \end{split}
\end{equation}
where the linear term is dropped since $\mathbf{u}^{k+1}$ and $\mathbf{w}^{k+1}$ are the minimizers of the problem \eqref{Relax u_update}, that is
$$\nabla_\mathbf{u} {L}_{\rho}(\mathbf{x}^{k+1},\mathbf{c}^{k+1},\mathbf{u}^{k+1},\mathbf{w}^{k+1},\mathbf{y}_1^k,\mathbf{y}_2^k) = 0,$$
$$\nabla_\mathbf{w} {L}_{\rho}(\mathbf{x}^{k+1},\mathbf{c}^{k+1},\mathbf{u}^{k+1},\mathbf{w}^{k+1},\mathbf{y}_1^k,\mathbf{y}_2^k) = 0.$$
Furthermore, according to \eqref{Relax y1_update} and \eqref{Relax y2_update}, we have
\begin{equation}\label{df_y1}
  \begin{split}
  &L_{\rho}(\mathbf{c}^{k+1},\mathbf{x}^{k+1},\mathbf{u}^{k+1},\mathbf{w}^{k+1},\mathbf{y}_1^{k},\mathbf{y}_2^k)
  \\
  &\hspace{1cm} -L_{\rho}(\mathbf{c}^{k+1},\mathbf{x}^{k+1},\mathbf{u}^{k+1},\mathbf{w}^{k+1},\mathbf{y}_1^{k+1},\mathbf{y}_2^k)\\
  =&-\frac{1}{\rho}\|\mathbf{y}_1^{k+1}-\mathbf{y}_1^k\|_2^2,
  \end{split}
\end{equation}
and
\vspace{-3pt}
\begin{equation}\label{df_y2}
 \begin{split}
  &L_{\rho}(\mathbf{c}^{k+1},\mathbf{x}^{k+1},\mathbf{u}^{k+1},\mathbf{w}^{k+1},\mathbf{y}_1^{k+1},\mathbf{y}_2^k) \\
  &\hspace{1cm} -L_{\rho}(\mathbf{c}^{k+1},\mathbf{x}^{k+1},\mathbf{u}^{k+1},\mathbf{w}^{k+1},\mathbf{y}_1^{k+1},\mathbf{y}_2^{k+1})\\
  =&-\frac{1}{\rho}\|\mathbf{y}_2^{k+1}-\mathbf{y}_2^k\|_2^2.
  \end{split}
\end{equation}
Adding both sides of \eqref{df_c}--\eqref{df_y2}, we can obtain
\begin{equation}\label{suff decrease}
  \begin{split}
    &L_{\rho}(\mathbf{c}^k,\mathbf{x}^k,\mathbf{u}^k,\mathbf{w}^k,\mathbf{y}_1^k,\mathbf{y}_2^k) \\
     &\hspace{1cm} -L_{\rho}(\mathbf{c}^{k+1},\mathbf{x}^{k+1},\mathbf{u}^{k+1},\mathbf{w}^{k+1},\mathbf{y}_1^{k+1},\mathbf{y}_2^{k+1}) \\ &\geq \frac{1}{2}\begin{bmatrix} \mathbf{u}^{k}-\mathbf{u}^{k+1}\\ \mathbf{w}^k-\mathbf{w}^{k+1} \end{bmatrix}^H
  \begin{bmatrix} \tilde{\rho}+\rho & \ -\tilde{\rho}\\ -\tilde{\rho} & \tilde{\rho}+\rho\\
\end{bmatrix}\begin{bmatrix} \mathbf{u}^{k}-\mathbf{u}^{k+1}\\ \mathbf{w}^k-\mathbf{w}^{k+1} \end{bmatrix}\\
     &-\frac{1}{\rho}(\|\mathbf{y}_1^{k+1}-\mathbf{y}_1^k\|_2^2+\|\mathbf{y}_2^{k+1}-\mathbf{y}_2^k\|_2^2).
  \end{split}
\end{equation}
Setting the gradient of the objective function in \eqref{Relax u_update} with respect to $\mathbf{u}$ as zero, we have the following derivations
\[
  \begin{split}
    0=&\nabla_{\mathbf{u}}L_{\rho}(\mathbf{c}^{k+1},\mathbf{x}^{k+1}, \mathbf{u}^{k+1},\mathbf{w}^{k+1},\mathbf{y}_1^k,\mathbf{y}_2^k) \\
     = & -\mathbf{y}_1^k+\tilde{\rho}(\mathbf{u}^{k+1}-\mathbf{w}^{k+1})-\rho(\mathbf{Ac}^{k+1}-\mathbf{u}^{k+1}) \\
     = & -\mathbf{y}_1^k+\tilde{\rho}(\mathbf{u}^{k+1}-\mathbf{w}^{k+1})+(\mathbf{y}_1^{k}-\mathbf{y}_1^{k+1}) \\
     = & \tilde{\rho}(\mathbf{u}^{k+1}-\mathbf{w}^{k+1}) - \mathbf{y}_1^{k+1},
  \end{split}
\]
then we can get
\vspace{-3pt}
\begin{equation}\label{y1solution}
\mathbf{y}_1^{k+1} = \tilde{\rho}(\mathbf{u}^{k+1}-\mathbf{w}^{k+1}),
\end{equation}
\vspace{-3pt}
 where the third equality comes from \eqref{Relax y1_update}. So, we have
\begin{equation}\label{dy1}
  \begin{split}
  &\lVert\mathbf{y}_1^{k+1}-\mathbf{y}_1^{k}\rVert_2^2 = \lVert\tilde{\rho}(\mathbf{u}^{k+1}-\mathbf{w}^{k+1})-\tilde{\rho}(\mathbf{u}^{k}-\mathbf{w}^{k})\rVert_2^2 \\
  &= \tilde{\rho}^2\lVert(\mathbf{u}^{k}-\mathbf{u}^{k+1})-(\mathbf{w}^{k}-\mathbf{w}^{k+1})\rVert_2^2 \\
  &= \begin{bmatrix} \mathbf{u}^{k}-\mathbf{u}^{k+1}\\ \mathbf{w}^k-\mathbf{w}^{k+1} \end{bmatrix}^H
  \begin{bmatrix} \tilde{\rho}^2 & \ -\tilde{\rho}^2\\ -\tilde{\rho}^2 & \tilde{\rho}^2\\
\end{bmatrix}\begin{bmatrix} \mathbf{u}^{k}-\mathbf{u}^{k+1}\\ \mathbf{w}^k-\mathbf{w}^{k+1} \end{bmatrix}.
  \end{split}
\end{equation}
Through similar derivations for the gradient of the problem \eqref{Relax u_update} with respect to $\mathbf{w}$, we can obtain \eqref{y2solution} and \eqref{dy2}.
\vspace{-3pt}
\begin{equation}\label{y2solution}
\mathbf{y}_2^{k+1} = -\tilde{\rho}(\mathbf{u}^{k+1}-\mathbf{w}^{k+1}),
\end{equation}
\begin{equation}\label{dy2}
\begin{split}
  &\lVert\mathbf{y}_2^{k+1}-\mathbf{y}_2^{k}\rVert_2^2=\lVert\tilde{\rho}(\mathbf{u}^{k}-\mathbf{w}^{k})-\tilde{\rho}(\mathbf{u}^{k+1}-\mathbf{w}^{k+1})\rVert_2^2 \\
  &=\begin{bmatrix} \mathbf{u}^{k}-\mathbf{u}^{k+1}\\ \mathbf{w}^k-\mathbf{w}^{k+1} \end{bmatrix}^H
  \begin{bmatrix} \tilde{\rho}^2 & \ -\tilde{\rho}^2\\ -\tilde{\rho}^2 & \tilde{\rho}^2\\
\end{bmatrix}\begin{bmatrix} \mathbf{u}^{k}-\mathbf{u}^{k+1}\\ \mathbf{w}^k-\mathbf{w}^{k+1} \end{bmatrix}.
\end{split}
\end{equation}
Plugging \eqref{dy1} and \eqref{dy2} into \eqref{suff decrease}, we obtain
\begin{equation}\label{dL1}
  \begin{split}
    &L_{\rho}(\mathbf{c}^k,\mathbf{x}^k,\mathbf{u}^k,\mathbf{w}^k,\mathbf{y}_1^k,\mathbf{y}_2^k)\\
     &\hspace{1cm}-L_{\rho}(\mathbf{c}^{k+1},\mathbf{x}^{k+1},\mathbf{u}^{k+1},\mathbf{w}^{k+1},\mathbf{y}_1^{k+1},\mathbf{y}_2^{k+1}) \\  \geq & \begin{bmatrix} \mathbf{u}^{k}-\mathbf{u}^{k+1}\\ \mathbf{w}^k-\mathbf{w}^{k+1} \end{bmatrix}^H
 \boldsymbol Q\begin{bmatrix} \mathbf{u}^{k}-\mathbf{u}^{k+1}\\ \mathbf{w}^k-\mathbf{w}^{k+1} \end{bmatrix},
  \end{split}
\end{equation}
where $\boldsymbol Q=\begin{bmatrix} \frac{\tilde\rho+\rho}{2}-\frac{2\tilde{\rho}^2}{\rho} & \ \frac{2\tilde{\rho}^2}{\rho}-\frac{\tilde{\rho}}{2}\\ \frac{2\tilde{\rho}^2}{\rho}-\frac{\tilde{\rho}}{2} & \frac{\tilde\rho+\rho}{2}-\frac{2\tilde{\rho}^2}{\rho}\\
\end{bmatrix}$ and its eigenvalues $\lambda(\mathbf{Q})$ are $\frac{\rho}{2}$ and $\frac{\rho^2+2\rho\tilde\rho-8\tilde\rho^2}{2\rho}$ respectively. We can verify that, when $\rho>2\tilde{\rho}>0$, the matrix $\mathbf{Q}$ is positive definite. Then, \eqref{dL1} can be simplified as
{
\begin{equation}\label{dL}
    \begin{split}
    &L_{\rho}(\mathbf{c}^k,\mathbf{x}^k,\mathbf{u}^k,\mathbf{w}^k,\mathbf{y}_1^k,\mathbf{y}_2^k)\\
     &\hspace{1cm}-L_{\rho}(\mathbf{c}^{k+1},\mathbf{x}^{k+1},\mathbf{u}^{k+1},\mathbf{w}^{k+1},\mathbf{y}_1^{k+1},\mathbf{y}_2^{k+1}) \\  \geq & \lambda_{\min}(\mathbf{Q})(\|\mathbf{u}^{k+1}-\mathbf{u}^{k}\|_2^2+\|\mathbf{w}^{k+1}-\mathbf{w}^{k}\|_2^2). \end{split}
\end{equation}}
Adding both sides of the above inequality from $k=1,2,...$, we can get
{
\begin{equation}\label{sum_df}
  \begin{split}
    &\hspace{-0.2cm}L_{\rho}(\mathbf{c}^1,\mathbf{x}^1,\mathbf{u}^1,\mathbf{w}^1,\mathbf{y}_1^1,\mathbf{y}_2^1)\\
     &\hspace{0.2cm}-\underset{k\rightarrow+\infty}\lim{L_{\rho}}(\mathbf{c}^{k+1},\mathbf{x}^{k+1},\mathbf{u}^{k+1},\mathbf{w}^{k+1},\mathbf{y}_1^{k+1},\mathbf{y}_2^{k+1}) \\ &\hspace{-0.2cm} \!\geq\! \lambda_{\min}(\mathbf{Q})\!\bigg(\!\displaystyle\sum_{k=1}^{+\infty}\!\|\mathbf{u}^{k+1}\!-\!\mathbf{u}^k\|_2^2\!
     +\!\displaystyle\sum_{k=1}^{+\infty}\!\|\mathbf{w}^{k+1}\!-\!\mathbf{w}^k\|_2^2\!\bigg)\!>\!0. \end{split}
\end{equation}}
Moreover, plugging \eqref{y1solution} and \eqref{y2solution} into the augmented Lagrangian function $L_{\rho}(\mathbf{c}^{k+1},\mathbf{x}^{k+1},\mathbf{u}^{k+1},\mathbf{w}^{k+1},\mathbf{y}_1^{k+1},\mathbf{y}_2^{k+1})$, we can derived it as
\begin{equation}\label{diff_Lk}
\begin{split}
&L_{\rho}(\mathbf{c}^{k+1},\mathbf{x}^{k+1},\mathbf{u}^{k+1},\mathbf{w}^{k+1},\mathbf{y}_1^{k+1},\mathbf{y}_2^{k+1}) \\
=&\frac{1}{2}\|\mathbf{S}_{\rm D}(\mathbf{c}^{k+1}\!-\!\mathbf{c}_{\rm o})\|_2^2\!+\!\tilde{\rho}\|\!\mathbf{Ac}^{k+1}\!-\!\frac{1}{2}\mathbf{u}^{k+1}\!-\!\frac{1}{2}\mathbf{w}^{k+1}\!\|_2^2\\
+&(\frac{\rho}{2}-\tilde{\rho})\big(\|\mathbf{Ac}^{k+1}-\mathbf{u}^{k+1}\|_2^2+\|\mathbf{x}^{k+1}-\mathbf{w}^{k+1}\|_2^2\big)\\
+&\tilde{\rho}\|\mathbf{x}^{k+1}-\frac{1}{2}\mathbf{u}^{k+1}-\frac{1}{2}\mathbf{w}^{k+1}\|_2^2.
\end{split}
\end{equation}
Since $\rho>2\tilde{\rho}$, we see that
\begin{equation}\label{Lk_lowbound}
L_{\rho}(\mathbf{c}^{k+1},\mathbf{x}^{k+1},\mathbf{u}^{k+1},\mathbf{w}^{k+1},\mathbf{y}_1^{k+1},\mathbf{y}_2^{k+1})\geq 0,\ \ \forall k.
\end{equation}
We can conclude from \eqref{sum_df} and \eqref{Lk_lowbound} that
\begin{subequations}\label{uw_gap}
  \begin{align}
  &\underset{k\rightarrow+\infty}\lim\mathbf{u}^{k+1}-\mathbf{u}^k=0,\\ &\underset{k\rightarrow+\infty}\lim\mathbf{w}^{k+1}-\mathbf{w}^k=0.
  \end{align}
\end{subequations}
Plugging \eqref{uw_gap} into \eqref{dy1} and \eqref{dy2} respectively, we get
\begin{subequations}\label{y_gap}
  \begin{align}
  &\underset{k\rightarrow+\infty}\lim\mathbf{y}_1^{k+1}-\mathbf{y}_1^k=0, \\ &\underset{k\rightarrow+\infty}\lim\mathbf{y}_2^{k+1}-\mathbf{y}_2^k=0.
  \end{align}
\end{subequations}
Combining the above results with \eqref{Relax y1_update} and \eqref{Relax y2_update}, we derive the following equalities
{
\begin{subequations}\label{dual_gap}
  \begin{align}
    &\underset{k\rightarrow+\infty}\lim\mathbf{Ac}^{k+1}-\mathbf{u}^{k+1}=0, \label{dual_gap_a}\\ &\underset{k\rightarrow+\infty}\lim\mathbf{x}^{k+1}-\mathbf{w}^{k+1}=0.\label{dual_gap_b}
  \end{align}
\end{subequations}}
\vspace{-3pt}

Next, let us show that $\mathbf{c}^{k+1}, \mathbf{x}^{k+1}, \mathbf{u}^{k+1}, \mathbf{w}^{k+1}, \mathbf{y}_1^{k+1}$ and $\mathbf{y}_2^{k+1}$ are bounded as $k\rightarrow+\infty$.

Plugging the limitation results \eqref{uw_gap} and \eqref{dual_gap} into \eqref{diff_Lk}, we can derive
 ${L_{\rho}}(\mathbf{c}^{k+1},\mathbf{x}^{k+1},\mathbf{u}^{k+1},\mathbf{w}^{k+1},\mathbf{y}_1^{k+1},\mathbf{y}_2^{k+1})$ as
\begin{equation}\label{diff_1infty}
 \begin{split}
  &\underset{k\rightarrow+\infty}\lim{L_{\rho}}(\mathbf{c}^{k+1},\mathbf{x}^{k+1},\mathbf{u}^{k+1},\mathbf{w}^{k+1},\mathbf{y}_1^{k+1},\mathbf{y}_2^{k+1}) \\
 &\hspace{-0.3cm}\! = \!\underset{k\rightarrow+\infty}\lim\frac{1}{2}\|\mathbf{S}_{\rm D}(\mathbf{c}^{k+1}\!-\!\mathbf{c}_{\rm o})\|_2^2\!+\! \underset{k\rightarrow+\infty}\lim \frac{\tilde\rho}{2}\|\mathbf{u}^{k+1}\!-\!\mathbf{w}^{k+1}\|_2^2 \\
 &\hspace{-0.3cm} \leq   {L_{\rho}}(\mathbf{c}^{1},\mathbf{x}^{1},\mathbf{u}^{1},\mathbf{w}^1,\mathbf{y}_1^{1},\mathbf{y}_2^{1}),
 \end{split}
\end{equation}
which means that $\|\mathbf{S}_{\rm D}\mathbf{c}^{k+1}\|_2^2$ and $\|\mathbf{u}^{k+1}-\mathbf{w}^{k+1}\|_2^2 $ are bounded as $k\rightarrow+\infty$. Moreover, since $\mathbf{c}^{k+1}$ satisfies the constraint \eqref{ori model_c}, we can conclude that $\|\mathbf{S}_{\rm F}\mathbf{c}^{k+1}\|_2^2$ is also bounded. Since $\mathbf{S}_{\rm D}+\mathbf{S}_{\rm F}$ is an identity matrix, we get that $\|\mathbf{c}^{k+1}\|_2$ is bounded as $k\rightarrow+\infty$. Plugging this result into \eqref{dual_gap_a}, we can see that $\|\mathbf{u}^{k+1}\|_2$ is bounded as $k\rightarrow+\infty$, which leads to $\|\mathbf{w}^{k+1}\|_2$ is also bounded. So, we can get that $\|\mathbf{x}^{k+1}\|_2$ is bounded from \eqref{dual_gap_b}. Furthermore, since $\mathbf{y}_1^{k+1} = \tilde{\rho}(\mathbf{u}^{k+1}-\mathbf{w}^{k+1})$ and $\mathbf{y}_2^{k+1} = -\tilde{\rho}(\mathbf{u}^{k+1}-\mathbf{w}^{k+1})$, we can conclude that the Lagrangian multipliers $\mathbf{y}_1^{k+1}$ and $\mathbf{y}_2^{k+1}$ are also bounded.

Combining the above bounded results with \eqref{uw_gap} and \eqref{y_gap}, we can obtain the following results
  \begin{equation}\label{converge uwy}
   \begin{split}
   &\underset{k\rightarrow+\infty}\lim\mathbf{u}^k=\mathbf{u}^*, \ \ \ \underset{k\rightarrow+\infty}\lim\mathbf{w}^k=\mathbf{w}^*, \\
   & \underset{k\rightarrow+\infty}\lim\mathbf{y}_1^k=\mathbf{y}_1^*, \ \ \ \underset{k\rightarrow+\infty}\lim\mathbf{y}_2^k=\mathbf{y}_2^*,\ \ \ \mathbf{y}_1^*=-\mathbf{y}_2^*.
   \end{split}
  \end{equation}
Since $\mathbf{u}^k$ and $\mathbf{w}^k$ are convergent as $k\rightarrow+\infty$, we have
$\underset{k\rightarrow+\infty}\lim\mathbf{Ac}^k=\mathbf{u}^*$ and $\underset{k\rightarrow+\infty}\lim\mathbf{x}^k=\mathbf{w}^*$.
Moreover, since $\mathbf{A}$ is full rank in columns, it means that
$\underset{k\rightarrow+\infty}\lim\mathbf{c}^k={\ell N}\mathbf{A}^H\mathbf{u}^*$. We can further obtain
\begin{equation}\label{converge cx}
   \begin{split}
   &\underset{k\rightarrow+\infty}\lim\mathbf{c}^k=\mathbf{c}^*, \ \ \ \underset{k\rightarrow+\infty}\lim\mathbf{x}^k=\mathbf{x}^*, \\
   &\ \  \mathbf{Ac}^*=\mathbf{u}^*, \ \ \ \ \ \ \ \ \mathbf{x}^*=\mathbf{w}^*,
   \end{split}
  \end{equation}
which concludes the proof of the first part of Theorem \ref{Relax convergence}.

Next, we consider to prove the second part of Theorem \ref{Relax convergence} that $(\mathbf{c}^*,\mathbf{x}^*,\mathbf{u}^*,\mathbf{w}^*)$ is a KKT point of the model \eqref{relax model}. Its proof is similar to the proof presented in Appendix B.
 Here, we denote $\tilde{L}(\mathbf{c},\mathbf{x},\mathbf{u},\mathbf{w},\mathbf{y}_1, \mathbf{y}_2, \mu, \gamma)$, $\tilde{L}^{\mathbf{c}}(\mathbf{c}, \mathbf{y}_1, \mu)$, $\tilde{L}^{\mathbf{x}}(\mathbf{x}, \mathbf{y}_2, \gamma)$, $\tilde{L}^{\mathbf{u}}(\mathbf{u}, \mathbf{w}, \mathbf{y}_1)$ and $\tilde{L}^{\mathbf{w}}(\mathbf{u}, \mathbf{w}, \mathbf{y}_2)$ as the Lagrangian functions of the problems \eqref{relax model}, \eqref{Relax c_update}, \eqref{Relax x_update}, and \eqref{Relax u_update} with respect to $\mathbf{u}$ and $\mathbf{w}$ respectively.
 $\mu$ and $\gamma$ are the Lagrangian multipliers corresponding to the constraints \eqref{relax model_c} and \eqref{relax model_b} respectively.
 When ADMM-Relax algorithm is convergent, we let $\mu^*$ and $\gamma^*$ denote the corresponding optimal Lagrangian multipliers.
 Since in every ADMM-Relax iteration $\mathbf{c}^{k+1}$ and $\mathbf{x}^{k+1}$ are always located in the feasible region, we can see that $\mathbf{c}^*$ and $\mathbf{x}^*$ satisfy the feasibility conditions, i.e.,
 \begin{equation}\label{relax cx pri feasibility}
   \mathbf{c^*\in\mathcal{C}}, \ \ \mathbf{x^*\in\mathcal{X}}.
 \end{equation}
 Since $\mu^{k*}\geq0$ in every ADMM-Relax iteration, it means
 \begin{equation}\label{relax dual feasibility}
    \mu^{*}\geq0.
 \end{equation}
 Moreover, from \eqref{Relax mu_c}, we see that the value of $\mu^{k*},\ \forall k,$ is nonzero or zero corresponding to the constraint \eqref{Relax c model_b}, which is active or inactive respectively. It means that $\mu^*$ satisfies the complementary slackness condition, i.e.,
 \begin{equation}\label{relax slackness}
   \mu^*(\|\mathbf{S}_{\rm F}\mathbf{c}^*\|_2^2-\beta\|\mathbf{S}_{\rm D}\mathbf{c}^*\|_2^2)=0.
 \end{equation}
 Furthermore, since $\mathbf{c}^{k+1}$, ${\mathbf{x}^{k+1}}$, ${\mathbf{u}^{k+1}}$, and ${\mathbf{w}^{k+1}}$ are the minimizers of the problems \eqref{Relax c_update}, \eqref{Relax x_update}, and \eqref{Relax u_update} respectively in the $k$th ADMM-Relax iteration, so they should satisfy
\begin{equation}\label{KKT cxuw 1}
  \begin{split}
  &\nabla_\mathbf{c} \tilde{L}^{\mathbf{c}}(\mathbf{c}^{k+1}, \mathbf{y}_1^{k}, \mu^{k+1})+\rho\mathbf{A}^H(\mathbf{Ac}^{k+1}-\mathbf{u}^k) = 0, \\
  &\nabla_\mathbf{x} \tilde{L}^{\mathbf{x}}(\mathbf{x}^{k+1}, \mathbf{y}_2^{k}, \gamma^{k+1})+\rho(\mathbf{x}^{k+1}-\mathbf{w}^k) = 0, \\
  &\nabla_\mathbf{u} \tilde{L}^{\mathbf{u}}(\mathbf{u}^{k+1}, \mathbf{w}^{k+1}, \mathbf{y}_1^{k})-\rho(\mathbf{Ac}^{k+1}-\mathbf{u}^{k+1}) = 0, \\
  &\nabla_\mathbf{w} \tilde{L}^{\mathbf{w}}(\mathbf{u}^{k+1}, \mathbf{w}^{k+1}, \mathbf{y}_2^{k})-\rho(\mathbf{x}^{k+1}-\mathbf{w}^{k+1}) = 0.
  \end{split}
\end{equation}
According to the convergence results \eqref{converge uwy} and \eqref{converge cx} , we can change \eqref{KKT cxuw 1} to \eqref{relax gradient zero} when $k\rightarrow+\infty$.
\begin{equation}\label{relax gradient zero}
  \begin{split}
  &\nabla_\mathbf{c} \tilde{L}^{\mathbf{c}}(\mathbf{c}^*, \mathbf{y}_1^*, \mu^*) = 0, \ \ \
  \nabla_\mathbf{x} \tilde{L}^{\mathbf{x}}(\mathbf{x}^*, \mathbf{y}_2^*, \gamma^*) = 0, \\
  &\nabla_\mathbf{u} \tilde{L}^{\mathbf{u}}(\mathbf{u}^*, \mathbf{w}^* \mathbf{y}_1^*)=0,\ \ \
  \nabla_\mathbf{w} \tilde{L}^{\mathbf{w}}(\mathbf{u}^*, \mathbf{w}^*, \mathbf{y}_2^*) = 0.
  \end{split}
\end{equation}
Since there are
\[
\begin{split}
&\nabla_\mathbf{c} \tilde{L}^{\mathbf{c}}(\mathbf{c}^*, \mathbf{y}_1^*, \mu^*)=\nabla_\mathbf{c} {\tilde{L}}(\mathbf{c}^*, \mathbf{x}^*, \mathbf{u}^*, \mathbf{w}^*, \mathbf{y}_1^*, \mathbf{y}_2^*),\\
&\nabla_\mathbf{x} \tilde{L}^{\mathbf{x}}(\mathbf{x}^*, \mathbf{y}_2^*, \gamma^*)=\nabla_\mathbf{x} {\tilde{L}}(\mathbf{c}^*, \mathbf{x}^*, \mathbf{u}^*, \mathbf{w}^*, \mathbf{y}_1^*, \mathbf{y}_2^*),\\
&\nabla_\mathbf{u} \tilde{L}^{\mathbf{u}}(\mathbf{u}^*, \mathbf{w}^* \mathbf{y}_1^*)=\nabla_\mathbf{u} {\tilde{L}}(\mathbf{c}^*, \mathbf{x}^*, \mathbf{u}^*, \mathbf{w}^*, \mathbf{y}_1^*, \mathbf{y}_2^*),\\
&\nabla_\mathbf{w} \tilde{L}^{\mathbf{w}}(\mathbf{u}^*, \mathbf{w}^*, \mathbf{y}_2^*)=\nabla_\mathbf{w} {\tilde{L}}(\mathbf{c}^*, \mathbf{x}^*, \mathbf{u}^*, \mathbf{w}^*, \mathbf{y}_1^*, \mathbf{y}_2^*),
\end{split}
\]
$(\mathbf{c}^*, \mathbf{x}^*, \mathbf{u}^*, \mathbf{w}^*)$ should also satisfy
\begin{equation}\label{relax stationary point}
\begin{split}
&\nabla_\mathbf{c} {\tilde{L}}(\mathbf{c}^*, \mathbf{x}^*, \mathbf{u}^*, \mathbf{w}^*, \mathbf{y}_1^*, \mathbf{y}_2^*)=0,\\
&\nabla_\mathbf{x} {\tilde{L}}(\mathbf{c}^*, \mathbf{x}^*, \mathbf{u}^*, \mathbf{w}^*, \mathbf{y}_1^*, \mathbf{y}_2^*)=0,\\
&\nabla_\mathbf{u} {\tilde{L}}(\mathbf{c}^*, \mathbf{x}^*, \mathbf{u}^*, \mathbf{w}^*, \mathbf{y}_1^*, \mathbf{y}_2^*)=0,\\
&\nabla_\mathbf{w} {\tilde{L}}(\mathbf{c}^*, \mathbf{x}^*, \mathbf{u}^*, \mathbf{w}^*, \mathbf{y}_1^*, \mathbf{y}_2^*)=0.
\end{split}
\end{equation}
Combining \eqref{relax cx pri feasibility}, \eqref{relax dual feasibility}, \eqref{relax slackness}, and \eqref{relax stationary point}, we can conclude that $(\mathbf{c}^*, \mathbf{x}^*, \mathbf{u}^*, \mathbf{w}^*)$ is some KKT point of the model \eqref{relax model}.

To prove the third part of Theorem \ref{Relax convergence}, we need to prove that $(\mathbf{c}^*,\mathbf{x}^*)$, combining the Lagrangian multipliers $\mathbf{y}^*$, $\mu^*$ and $\gamma^*$ satisfies the following KKT conditions
\begin{subequations}\label{quasi KKT}
    \begin{align}
      & \mathbf{c^*\in\mathcal{C}}, \ \       \mathbf{x^*\in\mathcal{X}}, \label{quasi KKT a} \\ &\underset{\tilde\rho\rightarrow+\infty}\lim\|\mathbf{Ac}^*-\mathbf{x}^*\|_2^2=0, \label{quasi KKT b} \\
      & \mu^*\geq0,  \label{quasi KKT c} \\
      & \mu^*(\|\mathbf{S}_{\rm F}\mathbf{c}^*\|_2^2-\beta\|\mathbf{S}_{\rm D}\mathbf{c}^*\|_2^2)=0,  \label{quasi KKT d} \\
      &
      \nabla_\mathbf{c} L(\mathbf{c}^*,\mathbf{x}^*,\gamma^*,\mu^*,\mathbf{y}^*)=0, \label{quasi KKT e} \\
      &\nabla_\mathbf{x} L(\mathbf{c}^*,\mathbf{x}^*,\gamma^*,\mu^*,\mathbf{y}^*)=0,  \label{quasi KKT f}
    \end{align}
\end{subequations}
where $\gamma$, $\mu$ and $\mathbf{y}$ are the Lagrangian multipliers corresponding to the constraints \eqref{ori model_b}, \eqref{ori model_c}, and \eqref{ori model_d} respectively, and $L(\mathbf{c},\mathbf{x},\gamma,\mu,\mathbf{y})$ is the Lagrangian function of the model \eqref{ori model}.
Notice here $\mathbf{y}^*=\mathbf{y}_1^*=-\mathbf{y}_2^*$.
The proof for $(\mathbf{c}^*,\mathbf{x}^*,\gamma^*,\mu^*,\mathbf{y}^*)$ satisfying \eqref{quasi KKT a}, \eqref{quasi KKT c}, \eqref{quasi KKT d} \eqref{quasi KKT e} , and \eqref{quasi KKT f} are the same as \eqref{relax cx pri feasibility}, \eqref{relax dual feasibility} and \eqref{relax slackness}.

Here, we only need to prove that $(\mathbf{c}^*,\mathbf{x}^*,\gamma^*,\mu^*, \mathbf{y}^*)$ also satisfies \eqref{quasi KKT b}.
 According to \eqref{diff_1infty}, we have
\begin{equation}\label{ineq_L}
 \begin{split}
   &L_{\rho}(\mathbf{c}^*,\mathbf{x}^*,\mathbf{u}^*,\mathbf{w}^*,\mathbf{y}_1^*,\mathbf{y}_2^*)\\
   =& \frac{1}{2}\|\mathbf{S}_{\rm D}(\mathbf{c}^*-\mathbf{c}_{\rm o})\|_2^2+\frac{\tilde\rho}{2}\|\mathbf{u}^*-\mathbf{w}^*\|_2^2 \\
   \leq & L_{\rho}(\mathbf{c}^1,\mathbf{x}^1,\mathbf{u}^1,\mathbf{w}^1,\mathbf{y}_1^1,\mathbf{y}_2^1).
 \end{split}
\end{equation}
If $\mathbf{c}^1\in\mathcal{C}$, $\mathbf{x}^1\in\mathcal{X}$, $\mathbf{Ac}^1=\mathbf{x}^1$ and $\mathbf{u}^1=\mathbf{w}^1$, then
\[
  L_{\rho}(\mathbf{c}^1,\mathbf{x}^1,\mathbf{u}^1,\mathbf{w}^1,\mathbf{y}_1^1,\mathbf{y}_2^1)=\frac{1}{2}\|\mathbf{S}_{\rm D}(\mathbf{c}^1-\mathbf{c}_{\rm o})\|_2^2.
\]
Then, we can change \eqref{ineq_L} to
\[
  \|\mathbf{u}^*-\mathbf{w}^*\|_2^2\leq\frac{1}{\tilde\rho}(\|\mathbf{S}_{\rm D}(\mathbf{c}^1-\mathbf{c}_{\rm o})\|_2^2-\|\mathbf{S}_{\rm D}(\mathbf{c}^*-\mathbf{c}_{\rm o})\|_2^2).
\]
Moreover, since $\mathbf{Ac}^*\!=\!\mathbf{u}^*$ and $\mathbf{x}^*\!=\!\mathbf{w}^*$, we can further get
\[
  \|\mathbf{Ac}^*-\mathbf{x}^*\|_2^2\leq\frac{1}{\tilde\rho}(\|\mathbf{S}_{\rm D}(\mathbf{c}^1-\mathbf{c}_{\rm o})\|_2^2-\|\mathbf{S}_{\rm D}(\mathbf{c}^*-\mathbf{c}_{\rm o})\|_2^2)=\mathcal{O}(\frac{1}{\tilde\rho}),
\]
which concludes the proof of the third part of Theorem \ref{Relax convergence}.

\section{Proof of Theorem \ref{iteration complexity}}
To be clear, here we rewrite \eqref{dL}
\[
  \begin{split}
    &\ \ \ L_{\rho}(\mathbf{c}^k,\mathbf{x}^k,\mathbf{u}^k,\mathbf{w}^k,\mathbf{y}_1^k,\mathbf{y}_2^k)\\
     &\hspace{1cm}-L_{\rho}(\mathbf{c}^{k+1},\mathbf{x}^{k+1},\mathbf{u}^{k+1},\mathbf{w}^{k+1},\mathbf{y}_1^{k+1},\mathbf{y}_2^{k+1}) \\ &\geq \lambda_{\min}(\mathbf{Q})(\|\mathbf{u}^{k+1}-\mathbf{u}^{k}\|_2^2+\|\mathbf{w}^{k+1}-\mathbf{w}^{k}\|_2^2).\\
  \end{split}
\]
 Summing both sides of the above inequality from $k=1,\dotsb, K$, we have
 \begin{equation}\label{dL rho}
   \begin{split}
     &L_{\rho}(\mathbf{c}^1,\mathbf{x}^1,\mathbf{u}^1,\mathbf{w}^1,\mathbf{y}_1^1,\mathbf{y}_2^1) \\
     &\hspace{0.5cm}-L_{\rho}(\mathbf{c}^{K+1},\mathbf{x}^{K+1},\mathbf{u}^{K+1},\mathbf{w}^{K+1},\mathbf{y}_1^{K+1},\mathbf{y}_2^{K+1}) \\ \geq& \lambda_{\min}(\mathbf{Q})\displaystyle\sum_{k=1}^K(\|\mathbf{u}^{k+1}-\mathbf{u}^k\|_2^2 +\|\mathbf{w}^{k+1}-\mathbf{w}^k\|_2^2).
   \end{split}
 \end{equation}
 Since $r = \underset{k}{\rm min}\{k|\|\mathbf{u}^{k+1}-\mathbf{u}^k\|_2^2 +\|\mathbf{w}^{k+1}-\mathbf{w}^k\|_2^2\leq\epsilon\}$, we can change \eqref{dL rho} to
 \begin{equation}\label{Relax rho}
   \begin{split}
     &L_{\rho}(\mathbf{c}^1,\mathbf{x}^1,\mathbf{u}^1,\mathbf{w}^1,\mathbf{y}_1^1,\mathbf{y}_2^1)\\
     &\hspace{1cm}-L_{\rho}(\mathbf{c}^{r+1},\mathbf{x}^{r+1},\mathbf{u}^{r+1},\mathbf{w}^{r+1},\mathbf{y}_1^{r+1},\mathbf{y}_2^{r+1}) \\ \geq& \lambda_{\min}(\mathbf{Q})r\epsilon.
   \end{split}
 \end{equation}
 Since we have $L_{\rho}(\mathbf{c}^{r+1},\mathbf{x}^{r+1},\mathbf{u}^{r+1},\mathbf{w}^{r+1},\mathbf{y}_1^{r+1},\mathbf{y}_2^{r+1})\geq L_{\rho}(\mathbf{c}^*,\mathbf{x}^*,\mathbf{u}^*,\mathbf{w}^*,\mathbf{y}_1^*,\mathbf{y}_2^*)$, \eqref{Relax rho} can be reduced to
 \[
   \begin{split}
     r &\leq \frac{1}{C\epsilon}\big(L_{\rho}(\mathbf{c}^1,\mathbf{x}^1,\mathbf{u}^1,\mathbf{w}^1,\mathbf{y}_1^1,\mathbf{y}_2^1)\\
     &\hspace{2cm}-L_{\rho}(\mathbf{c}^{*},\mathbf{x}^{*},\mathbf{u}^{*},\mathbf{w}^{*},\mathbf{y}_1^{*},\mathbf{y}_2^{*})\big),
   \end{split}
 \]
 where $C= \lambda_{\min}(\mathbf{Q})$, and $L_{\rho}(\mathbf{c}^{*},\mathbf{x}^{*},\mathbf{u}^{*},\mathbf{w}^{*},\mathbf{y}_1^{*},\mathbf{y}_2^{*})=\frac{1}{2}\|\mathbf{S}_{\rm D}(\mathbf{c}^*-\mathbf{c}_{\rm o})\|_2^2 + \frac{\tilde{\rho}}{2}\|\mathbf{u}^*-\mathbf{w}^*\|_2^2$, which concludes the proof of Theorem \ref{iteration complexity}.

\end{document}